\documentclass[twocolumn, floatfix]{aastex631}

\usepackage{ulem}
\usepackage{float}
\usepackage{gensymb}
\usepackage{array}
\usepackage{booktabs} 
\usepackage{multirow}

\usepackage{breqn}
\usepackage[flushmargin]{footmisc}

\usepackage{amsmath}

\shorttitle{}
\shortauthors{Vicentin et al.}

\graphicspath{{./}{figures/}}

\begin{document}

\title{The Journey to Dominance: How BCGs Evolve Differently from Other Massive Galaxies
}

\correspondingauthor{Vicentin, Marcelo C.}
\email{marcelo.vicentin@usp.br}

\author[0000-0002-9191-5972]{Marcelo C. Vicentin}
\affil{Universidade de S\~ao Paulo, Instituto de Astronomia, Geof\'isica e Ci\^encias Atmosf\'ericas, Departamento de Astronomia, \\ SP 05508-090, S\~ao Paulo, Brasil}
\affil{Department of Astrophysical Sciences, Princeton University, Peyton Hall, Princeton, NJ 08544, USA}

\author[0000-0002-0106-7755]{Michael A. Strauss}
\affil{Department of Astrophysical Sciences, Princeton University, Peyton Hall, Princeton, NJ 08544, USA}

\author[0000-0002-3876-268X]{Laerte Sodré Jr.}
\affil{Universidade de S\~ao Paulo, Instituto de Astronomia, Geof\'isica e Ci\^encias Atmosf\'ericas, Departamento de Astronomia, \\ SP 05508-090, S\~ao Paulo, Brasil}

\author[0000-0001-9320-4958]{Robert M. Yates}
\affil{Centre for Astrophysics Research, University of Hertfordshire, Hatfield, AL10 9AB, UK}

\author[0000-0003-2860-5717]{Pablo Araya-Araya}
\affil{Universidade de S\~ao Paulo, Instituto de Astronomia, Geof\'isica e Ci\^encias Atmosf\'ericas, Departamento de Astronomia, \\ SP 05508-090, S\~ao Paulo, Brasil}

\author[0000-0002-3281-9956]{Doris Stoppacher}
\affil{Departamento de Física Teórica, Módulo 15, Facultad de Ciencias, Universidad Autónoma de Madrid, Cantoblanco, 28049
Madrid, Spain}
\affil{Instituto de Astrofísica, Pontificia Universidad Católica de Chile, Campus San Joaquín, Avda. Vicuña Mackenna 4860, Santiago,
Chile}
\affil{Facultad de Físicas, Universidad de Sevilla, Campus de Reina Mercedes, Avda. Reina Mercedes s/n, 41012 Sevilla, Spain}

\begin{abstract}
We use the \texttt{L-GALAXIES} semi-analytic model to investigate the evolution of Brightest Cluster Galaxies (BCGs) found in clusters at $\rm z \sim 0$. BCGs are typically located in the central region of galaxy clusters, near the bottom of the potential well, exposing them to different environmental conditions compared to galaxies in the cluster outskirts or in the field. As a result, BCGs may follow a distinct evolutionary path and exhibit unique properties. We study the physical properties and merger histories of galaxies in 180 simulated clusters at $z \sim 0$, considering all cluster members with present-day stellar masses above $10^9 \ {\rm M_\odot}$ as the starting points for tracing their merger trees. We compare this sample of galaxies to a control sample of field galaxies and highlight their differences in evolution across cosmic time. We find that BCGs have distinct stellar mass formation histories compared to other massive galaxies from our control sample. Surprisingly, (proto)BCGs consistently become the most massive galaxy of their structure only at z $\sim$ 1.3. Despite this late dominance, (proto)BCGs are found to inhabit regions with higher galaxy and stellar mass density than the most massive galaxy in the structure throughout their entire history, indicating that their evolution is tightly linked to the environment from early times. These conditions shape a distinct evolutionary path for BCGs compared to other massive galaxies in clusters and in the field, underscoring the unique nature of BCGs.

\end{abstract}

\section{Introduction} \label{sec:intro}

Brightest Cluster Galaxies (BCGs) stand out from other galaxies even upon simple visual inspection. They are the largest and the most massive galaxy within a cluster, often recognized as the dominant galaxy. Their unique properties have intrigued researchers for decades, prompting numerous studies to better understand their characteristics. For instance, BCGs exhibit a narrow distribution of luminosities \citep{sandage72, schneider83b} and colors \citep{schneider83a, lauer14}. These distinct features have led to the use of BCGs as ``standard candles'' in cosmology \citep{gunn75, hoessel85, postman95}. 

The large size of BCGs is partly due to their extended stellar envelopes, which include material tidally stripped from satellite galaxies as they spiral toward the cluster core and merge with the BCG.
Such stripping and merging processes are expected to play a major role in shaping both the outer stellar halo and the intra-cluster light (ICL), although the relative contributions of these channels remain an active topic of research \citep{montes19, contini24}. Thus, this prominent stellar halo is closely linked to the fact that BCGs occupy a ``privileged'' location—typically at the center of the densest regions of the Universe, which coincide with the highest concentrations of dark matter. As a result, BCGs are usually found near the gravitational potential minimum of their clusters \citep{quintana82, jones84, lin04, lauer14, luo18, roche24}, as well as near the peak of X-ray emission from the Intra-Cluster Medium \citep[ICM;][]{rhee91, luo18}. Their large outer envelope (BCG+ICL) also accounts for their distinctive light profile, which differs from other giant elliptical galaxies \citep{oemler76,schombert86,huang18, huang22}. Furthermore, BCGs are aligned with the cluster's structure--the alignment of major axis of the BCG toward the distribution of cluster satellite galaxies--and with the large-scale structure in which the cluster is embedded \citep{sastry68, binggeli82, lambas88, niederste10, hao11, smith23}, and exhibit a small velocity offset relative to the cluster's mean redshift \citep{postman95, vandenbosch05,coziol09}.

Several studies suggest that the exceptional properties of BCGs cannot be explained solely as statistical extremes of the cluster galaxy population \citep[e.g.,][]{tremaine77, loh06, lin10, shen14}. For instance, \citet{dalal21}, using BCG and second-ranked cluster galaxy masses from the HSC-SSP Camira catalog \citep{oguri18}, found that the mass gap between these two populations up to $\rm z \sim 1$ is too large to be consistent with BCGs being statistical extremes of the mass distribution of other cluster member galaxies. This reinforces the notion that BCGs, cannot be regarded as mere statistical outliers but rather as a special class of galaxies shaped by distinct evolutionary processes--processes that ultimately shape them into the dominant galaxies of their host clusters.

Simulations play a pivotal role in understanding the expected evolution of galaxies over cosmic time. The pioneering work of \citet{delucia07} predicted, using semi-analytic models, that BCGs assembled half of their stellar mass as early as $z \sim 5$, earlier than suggested by most observational estimates at the time. Those authors already compared their predictions with available data and discussed the associated uncertainties. Subsequent improvements to galaxy formation models--including more detailed environmental processes--and the explicit modeling of the halo component \citep[e.g.,][]{monaco06, lin13, shankar15, henriques15, henriques20,contini24}--have led to substantially better agreement with observations of BCG stellar mass growth. These developments highlight the importance of accurately modeling the dense environments where the most massive halos reside, as they critically shape the evolution of their central galaxies.

The goal of this study is to investigate the stellar mass evolutionary history of BCGs and assess how their growth differs from other massive galaxies, both in clusters and in the field. Are BCGs intrinsically unique, or do they become distinct due to their environment? Do they follow fundamentally different evolutionary paths, and are the BCGs we observe today the same as those at high redshift, or do they emerge as the dominant galaxies over time? To address these questions, we analyze the stellar mass evolution of BCGs using semi-analytical galaxy formation models, comparing them with other massive galaxies to uncover what sets BCGs apart in their evolutionary trajectories.

This paper is organized as follows. In Section \ref{sec:data}, we briefly describe the Millennium \citep{springel05} simulation and the semi-analytical galaxy formation model \texttt{L-GALAXIES} \citep[e.g.,][]{white91, guo11, henriques15, henriques20}, which form the basis of our analysis, emphasizing important definitions and concepts that we use across this work. Section \ref{sec: SMH} introduces key definitions and concepts, and presents our analysis of the stellar mass evolution of (proto)BCGs in comparison to other cluster and field massive galaxies. In Section \ref{sec: dominance}, we explore how BCGs emerge as the dominant galaxy within their structures, focusing on their intrinsic stellar mass growth relative to the most massive companion galaxies as a function of redshift and the role of the environment in shaping this evolution. Finally, Section \ref{sec:conclusions} provides a summary of our findings and highlights the main conclusions of this study. For this study, we adopt the $\rm \Lambda$CDM cosmological model with $\rm h = 0.673$, $\rm \Omega_{\rm m} = 0.315$, and $\rm \Omega_{\Lambda} = 0.685$ \citep{planck1}.

\section{Data} \label{sec:data}


In this study, we use data from the application of the \citet{ayromlou21} (hereafter A21) version of the semi-analytic model for galaxy formation and evolution \texttt{L-GALAXIES} to the dark matter-only (DMO) Millennium Simulation \citep{springel05}. The simulation traces the history of 2160$^{3}$ dark matter particles with a mass of $\rm 9.6 \times 10^{8} \ h^{-1} \ M_{\odot}$ from $\rm z = 127$ to the present, within a box with a side length of $\rm 500 \ h^{-1} \ Mpc$.

To apply the semi-analytic models, A21 re-scaled the original Millennium Simulation using the cosmological parameters obtained by the \citet{planck1}, i.e., $\sigma_{8} = 0.829, \rm H_{0} = 67.3 \ km \ s^{-1} \ Mpc^{-1}, \Omega_{\Lambda} = 0.685, \Omega_{m} = 0.315, \Omega_{b} = 0.0487$, and $\rm n = 0.96$, following the method proposed by \citet{angulo10} and updated by \citet{angulo15}. With these parameters, the simulation box volume is $\rm 714^{3} \ Mpc$ and dark matter particles have a mass of 1.43 $\rm \times 10^{9} \ M_{\odot}$. 

In the following subsections, we briefly outline the key features of the data used in our analysis. 

\subsection{Merger trees}
\label{sec: merger_trees}

To construct the merger trees in the Millennium Simulation, the data were post-processed at each time step using a friends-of-friends (FOF) algorithm to create groups of particles that are within one-fifth of the mean inter-particle distance \citep{davis85}. Dark matter subhalos were then identified within each FOF group using the \texttt{SUBFIND} algorithm \citep{springel01}. Each subhalo in a given snapshot that contains 20 or more particles, i.e., $\rm M \geq 2.86 \times 10^{10} \ M_{\odot}$, is connected to a unique descendant in the subsequent snapshot. This linking of subhalos across time facilitates the construction of merger trees by accounting for the substructures within dark matter halos. The main halo within a given FOF group is then defined as the one with the highest mass.



\subsection{L-GALAXIES}
\label{sec: lgalaxies}

The \texttt{L-GALAXIES} semi-analytic model for galaxy formation and evolution has been continuously developed for more than 30 years to implement baryonic physical processes on top of merger trees in dark matter-only (DMO) N-body simulations through a set of coupled differential equations \citep[e.g.,][]{White89, white91, kauffmann99,springel05}. With the advent of the Millennium Simulation in 2005, it became possible to apply \texttt{L-GALAXIES} in an environment with sufficient resolution to resolve individual galaxies within a large cosmological volume \citep[e.g.,][]{delucia06, boylan09, yates13, henriques15}.

In \texttt{L-GALAXIES 2020} (\citealp{henriques20}, hereafter H20), baryonic physics is treated by introducing an expected fraction of primordial hot gas, based on the cosmic abundance of baryons ($\rm \Omega_{b}/\Omega_m$), into each dark matter subhalo when it first forms (i.e. when it is first resolved within the underlying N-body simulation). This gas is assumed to cool over time by radiating energy. The efficiency of the cooling process will determine when the cold gas will form stars within a given dark matter halo. In \texttt{L-GALAXIES 2020}, this cold gas is partitioned into atomic hydrogen (HI) and molecular hydrogen ($\rm H_{2}$). When molecular hydrogen reaches a certain surface density and metallicity threshold, it collapses and forms stars within galaxies. Newly formed stars and the cold gas are organized into a galactic disc, which is divided into 12 concentric, independently tracked rings. 

Stars can also be transferred to a stellar bulge by galaxy mergers or disc instabilities. During these events, part of the cold gas is funneled toward the galaxy center, where it may be converted into stars through merger-driven or instability-driven starbursts (see Section~\ref{sec: MBISH}). Feedback from supernovae and active galactic nuclei (AGN) can reheat or eject part of this gas, regulating subsequent star formation. In \texttt{L-GALAXIES~2020}, supernova feedback both reheats and ejects cold gas, whereas AGN feedback operates differently: instead of directly ejecting gas, AGNs reduce the cooling rate of the hot gas reservoir, mimicking the effect of energy deposition from radio jets that keep the circumgalactic medium hot and turbulent. This mechanism indirectly suppresses further cold-gas accretion and thus quenches star formation in massive galaxies. 

Bulges can form through major and minor mergers or via disk instabilities: major mergers transfer all progenitor stars to the bulge, minor mergers add stars from the smaller progenitor to the bulge while preserving the stellar disk of the larger one, and disk instabilities move material inward towards the galaxy center (Section~\ref{sec: DBH}).

Additionally, stars tidally stripped from satellites by the cluster potential are added to the stellar halo of the central galaxy, contributing to the intracluster light (ICL; see Section~\ref{sec: FH}). The model also includes environmental effects such as tidal and ram-pressure stripping, which directly affect a satellite galaxy’s gas reservoirs and morphology.\footnote{For readers interested in more details about the models implemented in the \texttt{L-GALAXIES 2020} framework, we refer to the supplementary material available at \url{https://lgalaxiespublicrelease.github.io/Hen20_doc.pdf}.} These environmental processes are crucial for realistically determining the evolution of galaxies in dense environments such as (proto)clusters, the focus of our study.

Given the importance of environmental effects on cluster member galaxies, we use the version of \texttt{L-GALAXIES 2020} described in A21. This version includes an improvement in the modeling of tidal and ram pressure stripping processes, which directly impact the evolution of galaxies.

In the same way that progenitor dark matter subhalo information is stored in the public Millennium database\footnote{\url{https://virgodb.dur.ac.uk/}} \citep{lemson06}, the results generated by \texttt{L-GALAXIES} also store the information of the progenitors of a given galaxy. This allows the tracking of the history of a galaxy, including all the objects that contribute to its formation history.

\subsection{Galaxy cluster merger trees \& definitions}
\label{sec: defs}

The aim of this work is to statistically analyze the evolution of BCGs compared to the second most massive cluster member ($\rm 2MM$) and massive field galaxies defined at $z \sim 0$. We use the definition of galaxy clusters in simulations similar to \citet{chiang13}, classifying structures into three types according to their FOF halo masses at $\rm z = 0$: 

\begin{itemize}
    \item[] \textit{Fornax-like}: $1.5 \times 10^{14} M_{\odot} \leq M_{DM} < 3 \times 10^{14} M_{\odot}$

    \item[] \textit{Virgo-like}: $3 \times 10^{14} M_{\odot} \leq M_{DM} < 10^{15} M_{\odot}$

    \item[] \textit{Coma-like}: $M_{DM} \geq 10^{15} M_{\odot}$.
        
\end{itemize}

The lower limit of $M_{\rm DM}=1.5\times10^{14}\,{\rm M_\odot}$ corresponds approximately to the transition between massive galaxy groups and bona fide clusters, where the ICM becomes virialized \citep[e.g.,][]{bleem20, hurier21}.

Cluster membership is defined through the FOF groups at $z \sim 0$, where we consider as \textit{Cluster Members} all galaxies associated with the main FOF halo with $M_{\rm DM} \ge 1.5\times10^{14} \ {\rm M_\odot}$, including the central and satellite galaxies (types 0, 1, and 2 in \texttt{L-GALAXIES}). At higher redshifts, the members of a given structure are defined as all progenitors of those galaxies identified as cluster members at $z \sim 0$, i.e., galaxies that belong to the merger trees of the $z \sim 0$ members. This approach ensures that the evolutionary history of each present-day cluster is consistently traced back in time without imposing any requirement on distance of the galaxy from the cluster progenitor.

Our sample includes all galaxies with $M_{\star} \ge 10^{9}\,{\rm M_\odot}$ at $z=0$ belonging to 180 randomly selected clusters, equally distributed among the three mass ranges. For each of these galaxies, we also include all progenitors traced through their merger trees. The division of structures by dark matter halo mass aims to identify potential physical differences in the histories of the galaxies residing within them. We are selecting all Coma-like structures (60) within the simulation volume at $z \sim 0$, as they are much rarer than the lower mass halos. The BCG (2MM) of each cluster is defined as the galaxy with the highest (second-highest) stellar mass within the most massive halo at $\rm z = 0$. As expected, BCGs identified in this way correspond to the central galaxy of the structure at $\rm z = 0$, i.e., a ``type 0'' galaxy according to the \texttt{L-GALAXIES} classification.

To understand the difference between field and cluster galaxies, we create a sample of $\rm \sim 5 \times 10^{3}$ randomly selected massive field galaxies at $z \sim 0$ with $\rm M_{\star} \geq 10^{11} \ M_{\odot}$ that inhabit halos with $\rm M_{DM} < 10^{13.8} \ M_{\odot}$, along with their progenitors. These galaxies are all central galaxies (type~0) in \texttt{L-GALAXIES}. No attempt was made to match their stellar mass distribution to that of the cluster populations. These field galaxies are typically less massive than BCGs,, reflecting the physical rarity of field centrals with comparable stellar masses. In our sample, BCGs have a median stellar mass of $\log(M_{\star}/{\rm M_\odot}) = 11.88^{+0.17}_{-0.19}$,  whereas field galaxies $\log(M_{\star}/{\rm M_\odot}) = 11.14^{+0.13}_{-0.08}$. This difference is expected, since only a small number of field centrals reach the extreme stellar masses of BCGs, but it should be kept in mind when comparing the evolutionary trends among the samples.

As we move to higher redshifts, the dark matter mass of the structure $\rm M_{DM}$ decreases and, at a certain point in its history, it falls below the threshold of our cluster definition $\rm M_{DM} \geq 1.5 \times 10^{14} \ M_{\odot}$. From this point onward, i.e., to higher redshift, the structure is classified as a \textit{galaxy protocluster} \citep[e.g.,][]{chiang13, araya21}. Consequently, the main progenitor of the BCG will be defined as the \textit{protoBCG}. Thus, a protocluster is a structure with $\rm M_{DM} < 1.5 \times 10^{14} \ M_{\odot}$ that will exceed this mass at some point in its future by z $\geq$ 0. In contrast, the selected field galaxies belong to halos that remain below this threshold at z = 0. This operational definition, while reasonable for simulations where the full evolutionary history of halos is accessible, is somewhat arbitrary and not identical to the observational criteria often adopted, where protoclusters are generally recognized as sufficiently overdense regions that are likely to collapse into clusters \citep{overzier16}.

\section{Stellar Mass Histories} \label{sec: SMH}

In this section, we study the stellar mass assembly histories of our selected BCGs. We adopt the same terminology of \citet{delucia07} (hereafter DLB07) for each galaxy:

\begin{itemize}
    \item[-] \textit{Identity redshift} ($z_{id}$): The redshift at which the galaxy's main progenitor undergoes its last major merger event, in which the accreted stellar mass is greater than one third of the main progenitor galaxy's stellar mass in the previous time step\footnote{In DLB07, they term this redshift as the \textit{Extended Identity Time}.}.

    \item[-]{\textit{Assembly history}: Physically, this traces when the main progenitor of the galaxy assembled its mass. We compute
    $M_{\star}^{\rm MainProg}(z)/M_{\star}(z\simeq 0)$, i.e. the stellar mass of the main-progenitor branch at redshift $z$,
    normalized by the present-day stellar mass. This quantity increases as the main branch grows via in situ star formation and
    mergers}.

    \item[-]{\textit{Star formation history}: Traces when the \emph{stars that will end up in the $z\simeq 0$ galaxy} were formed,
    irrespective of which progenitor formed them along its merger tree. We compute $\Big[\sum_{\rm progenitors} M_{\star}(z)\Big]/M_{\star}(z\simeq 0)$, i.e. the total stellar mass contained in \emph{all}
    progenitors at redshift $z$, normalized by the present-day stellar mass. This ratio can temporarily exceed unity because part of
    that stellar mass is later stripped/disrupted into the stellar halo/ICL or otherwise not retained in the bound galaxy at
    $z\simeq 0$}.

\end{itemize}

Other concepts and quantities we use in this paper are:

\begin{itemize}
    \item[-] \textit{Cluster formation redshift} ($z_{14}$): The redshift of the first snapshot at which the FOF group that hosts the main-branch halo of the $z\simeq0$ system reaches $M_{\rm DM}\ge 1.5\times10^{14}\,{\rm M_\odot}$. Operationally, we follow the main progenitor of the $z\simeq0$ halo backward in time and, at each snapshot, query the mass of its parent FOF group (FOF defined as in §2.1).

    \item[-] \textit{Half-final stellar mass assembled redshift} ($z_{50a}$): The redshift at which the main progenitor of a given galaxy reaches 50\% of its final stellar mass.

    \item[-] \textit{Half-final stellar mass formed redshift} ($z_{50f}$): The redshift at which the sum of the progenitors of a given galaxy reaches 50\% of its final stellar mass.

    \item[-] \textit{Stellar mass assembly rate}: The increase in the galaxy's stellar mass per unit time at each snapshot, i.e., $\Delta M_{\star}/\Delta t$.

    \item[-] \textit{Stellar mass assembly rate peak redshift} ($z_{peak, AR}$): The redshift at which the median stellar mass assembly rate of a given galaxy population (BCGs, 2MMs, or Field galaxies) reaches its maximum.
\end{itemize} 

Additionally, we will use other quantities related to stellar mass provided by \texttt{L-GALAXIES}, such as the amount of stellar mass in a galaxy built due to \textit{mergers} and starbursts, or secular \textit{in situ} star formation, which will be important for understanding the main mechanisms of stellar formation throughout the histories of the galaxies (Section \ref{sec: MBISH}). We will also consider the \textit{bulge} and \textit{disk} mass separately to understand how these components, and thus the morphology of the galaxy, evolve (Section \ref{sec: DBH}).

\subsection{Assembly history}
\label{sec: AH}

\begin{figure*}

	\includegraphics[width=\textwidth]{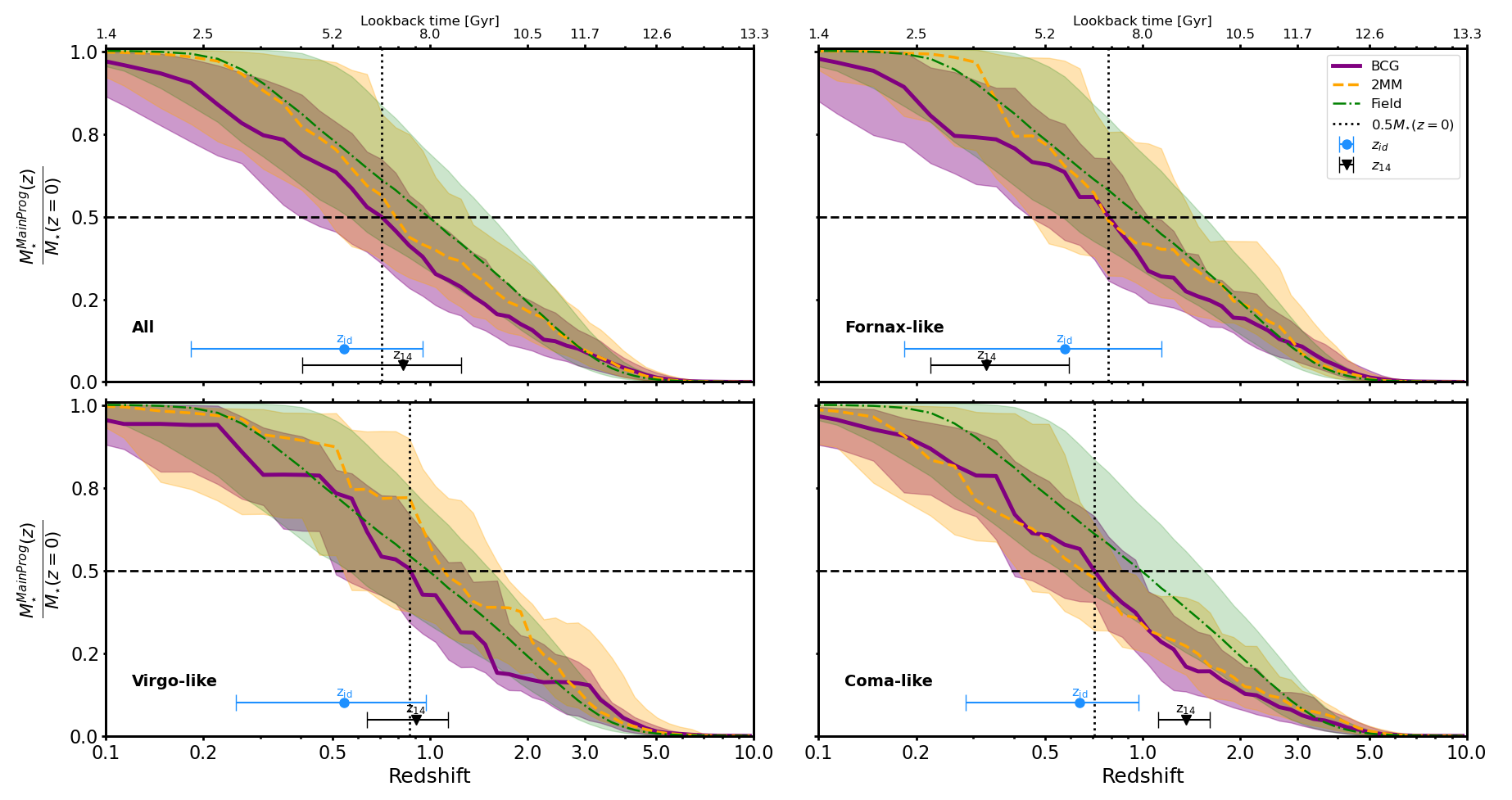}
    \caption{Stellar mass assembly histories as a function of redshift (lower axis) and lookback time (upper axis). The figure presents plots of the assembly histories of all structures (top-left) and of different types of structures: Fornax-like (top-right), Virgo-like (bottom-left), and Coma-like (bottom-right), as defined in Section \ref{sec: defs}. The solid purple, dashed orange, and dot-dashed green curves in each panel denote the median assembly history of BCGs, the second most massive galaxy in the structure ($\rm 2MM$), and massive field galaxies, respectively. The black inverted triangle with error bars, cyan dot with error bars, and dotted line indicate the median and inter-quartile range of the cluster formation redshift ($z_{14}$), the median and inter-quartile range of the identity redshift ($z_{id}$), and the median half-final stellar mass assembled of BCGs ($z_{50a}$), respectively. The dashed horizontal line marks the threshold where the median BCG reaches 50\% of its final stellar mass. The shaded areas, following the same color schemes, encompass the 25th and 75th percentiles over the galaxies in each subsample.     
}
    \label{fig: sm_hist}
\end{figure*}

Figure \ref{fig: sm_hist} shows the median stellar-mass assembly histories for BCGs, the 2MMs, and massive field galaxies. Additionally, the plots display other important statistics in the history of BCGs, such as $z_{id}$, $z_{14}$, and $z_{50a}$. To understand the influence of the dark matter halo mass in which galaxies reside on their stellar assembly histories, we show results for all 180 structures (top-left panel), and for only Fornax-like (top-right), Virgo-like (bottom-left), and Coma-like (bottom-right), following the definitions presented in Section \ref{sec: defs}.

Although the assembly histories of BCGs, $\rm 2MMs$, and massive field galaxies, overlap within the inter-quartile range, the median trends reveal a systematic difference. Specifically, massive field galaxies and $\rm 2MMs$ tend to assemble their stellar mass more rapidly, reaching their final mass around $z \sim 0.2$, whereas BCGs continue assembling stellar mass down to $z \sim 0$. 

This difference between the BCG and $\rm 2MM$ curve can be attributed to the fact that, in a cluster environment with its high velocity dispersion, galaxies experience significantly fewer merger events, effectively halting their mass growth through this process \citep[e.g.,][]{ostriker80, deger18}. By $z \sim 0.2$, nearly all the structures in our sample have already reached the cluster mass threshold ($z_{14}$) and had time to dynamically mature, further limiting merger-driven growth (see also Section \ref{sec: MBISH}). Additionally, environmental effects strip the hot gas reservoir, preventing further cooling and progressively quenching star formation. BCGs are exceptions, as they continue to accrete other galaxies that move towards the cluster core due to dynamical friction with the cluster environment and are eventually cannibalized by the central galaxy \citep[][]{ostriker75, nipoti03}. 

\begin{figure*}[ht]

	\includegraphics[width=\textwidth]{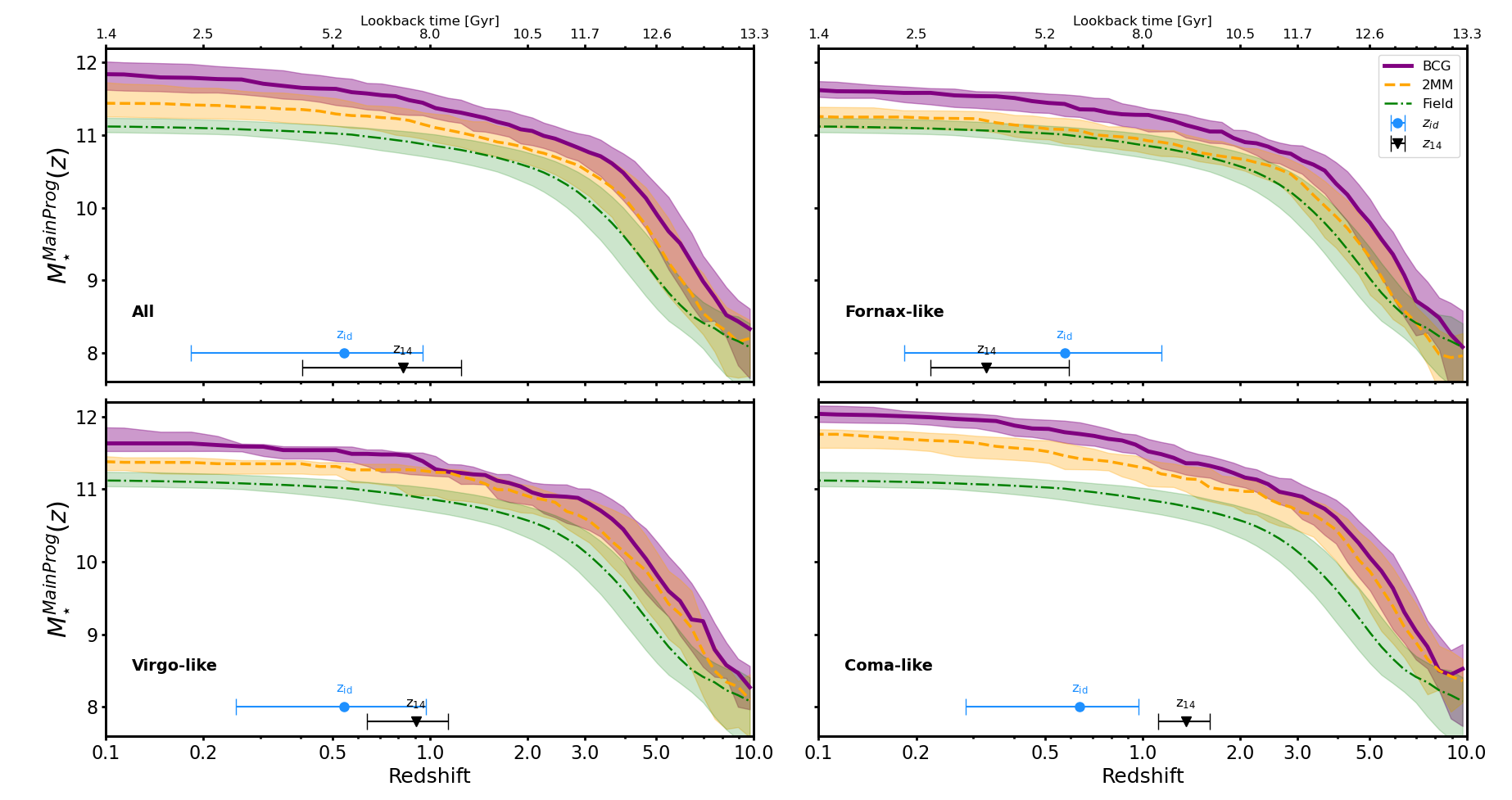}
    \caption{Stellar mass absolute assembly histories as a function of redshift (lower axis) and lookback time (upper axis). The panels, curves and colors follow the same scheme as in Figure \ref{fig: sm_hist}. 
}
    \label{fig: sm_abs_assembly}
\end{figure*}

Massive field galaxies at $\rm z \sim 0.2$ are already more isolated, and therefore, merger events are rare. At these redshifts, almost all of these massive galaxies, regardless of the environment, are quiescent \citep[having undergone mass quenching; ][]{peng10}, i.e., they have little or no \textit{in situ} star formation. Even though BCGs assemble their final mass may be more delayed than other massive galaxies, they assemble more mass than any other type of galaxy, as shown in Figure \ref{fig: sm_abs_assembly}. (Proto)BCGs and $\rm 2MMs$ residing in the most massive structures (Coma-like) are systematically more massive than their counterparts in lower-mass structures, exhibiting an even larger discrepancy when compared to massive field galaxies.

Table \ref{tab: assembly} shows the median and percentile ranges of the half-final stellar mass assembly ($ z_{50a}$), identity ($ z_{id}$), and cluster formation ($z_{14}$) redshifts. For the full sample of 180 clusters, the assembly sequence progresses as follows: the cluster formation redshift occurs at $z_{14} \sim 0.8$, the BCG's main progenitor accumulates half of its final mass by $z \sim 0.7$, and undergoes its last major merger at $z_{id} \sim 0.5$. The order of these events varies depending on the type of structure. By construction, in less massive Fornax-like clusters, the BCG has already acquired more than half of its final mass and its identity before/during the transition from protocluster to cluster. In contrast, for the more massive clusters, Coma-like, the structures already have $\rm M_{DM} \geq 1.5 \times 10^{14} \ M_{\odot}$ when the BCGs acquire their identity and 50\% of their final mass. Furthermore, BCGs assemble most of their final mass not as protoBCGs, but after residing in a structure with $\rm M_{DM} > 1.5 \times 10^{14} \ M_{\odot}$, i.e., a galaxy cluster. This occurs in up to $\sim$ 75\% and nearly 100\% of the cases for Virgo- and Coma-like structures, respectively. The results found here for $z_{50a}$ and $z_{id}$ are consistent with those reported by DLB07.

Regarding the formation redshift of galaxy clusters, \citet{huang20} present in their Figure 4 a compilation of clusters with masses above $\rm 10^{14} \ M_{\odot}$ identified through the Sunyaev-Zeldovich (SZ) effect across various surveys. The plot shows that identifications start at around $\rm z \sim 1.6$ with a few examples, and as we move to lower redshifts, the number of these structures increases significantly. These observations are therefore consistent with our $\rm z_{14}$ results. Figure 4 of \citet{overzier16} also showcases a collection of clusters and protoclusters identified up to the publication date of that review. There are five examples of high-z clusters within $\rm 1.6 < z < 2$, and at higher redshifts, only protoclusters are noted. These findings are also consistent with our results, where most of the more massive clusters transition from the protocluster phase to the cluster phase at $z_{14} = 1.4^{+0.3}_{-0.2}$. 

The stellar masses of BCGs obtained in this work are summarized in Table~\ref{tab:bcg_mass_comparison}, where we compare with the observations of \citet{erfanianfar19} and \citet{lin13}, for different redshifts and halo masses. \citet{erfanianfar19} analyzed 416 BCGs from the X-ray clusters in SPIDERS-CODEX survey catalog, obtaining stellar mass estimates for BCGs in halos of different masses across two redshift ranges. Our results show good agreement with their estimates, except for the most massive systems at $\rm 0.1 < z < 0.3$, where we obtain slightly higher values.

Similarly, \citet{lin13} studied BCGs in clusters from the Spitzer IRAC Shallow Cluster Survey \citep[ISCS; ][]{eisenhardt08}, covering a broad redshift range. Their estimates suggest an increase in stellar mass by a factor of $\sim2.3$ between $z = 1.5$ and 0.5, with a slower growth at $\rm z < 0.5$, aligning well with the results presented here at all redshifts.

We note that our analysis follows the progenitors of halos identified at $z \sim 0$, whereas the observational samples cited above consist of clusters identified independently at each redshift. Hence, these comparisons are qualitative, aiming to highlight similarities and differences between the evolutionary trends predicted by galaxy formation models and those observed.



\begin{table}[h!]
    \caption{Median redshifts and interquartile ranges for key evolutionary milestones.}
    \centering
    \renewcommand{\arraystretch}{1.5} 
    \begin{tabular}{c c c c c}
        \toprule
        & \textbf{All} & \textbf{Fornax-like} & \textbf{Virgo-like} & \textbf{Coma-like} \\
        \midrule
        \textbf{$z_{\rm{50a}}^{\rm{BCG}}$} & $0.7^{+0.3}_{-0.3}$ & $0.8^{+0.3}_{-0.3}$ & $0.6^{+0.2}_{-0.3}$ & $0.7^{+0.2}_{-0.3}$ \\
        
        \textbf{$z_{\rm{50a}}^{\rm{2MM}}$} & $0.8^{+0.6}_{-0.3}$ & $0.8^{+0.7}_{-0.3}$ & $0.9^{+0.7}_{-0.3}$ & $0.6^{+0.2}_{-0.2}$ \\
        
        \textbf{$z_{\rm{50a}}^{\rm{Field}}$} & $1.0^{+0.5}_{-0.4}$ & -- & -- & -- \\
        
        \textbf{$z_{\rm{peak, \ AR}}^{\rm{BCG}}$} & $3.8 \pm 1$ & $3.8 \pm 0.9$ & $3.8 \pm 1$ & $4.2 \pm 1.2$ \\
        
        \textbf{$z_{\rm{peak,\ AR}}^{\rm{2MM}}$} & $3.5 \pm 1$ & $2.8 \pm 0.8$ & $3.2 \pm 0.9$ & $4 \pm 1.1$ \\
        
        \textbf{$z_{\rm{peak, \ AR}}^{\rm{Field}}$} & $2.6 \pm 1$ & -- & -- & -- \\
        
        \textbf{$z_{\rm{50f}}^{\rm{BCG}}$} & $3.6^{+0.3}_{-0.3}$ & $3.4^{+0.3}_{-0.3}$ & $3.6^{+0.3}_{-0.3}$ & $4.0^{+0.3}_{-0.3}$ \\
        
        \textbf{$z_{\rm{50f}}^{\rm{2MM}}$} & $3.1^{+0.3}_{-0.7}$ & $2.6^{+0.7}_{-0.6}$ & $3.1^{+0.5}_{-0.7}$ & $3.1^{+0.5}_{-0.2}$ \\
        
        \textbf{$z_{\rm{50f}}^{\rm{Field}}$} & $2.2^{+0.4}_{-0.6}$ & -- & -- & -- \\    
        
        \textbf{$z_{\rm{id}}$} & $0.5^{+0.4}_{-0.4}$ & $0.6^{+0.6}_{-0.4}$ & $0.5^{+0.3}_{-0.2}$ & $0.6^{+0.3}_{-0.4}$ \\
        
        \textbf{$z_{14}$} & $0.8^{+0.4}_{-0.4}$ & $0.3^{+0.2}_{-0.1}$ & $0.9^{+0.2}_{-0.3}$ & $1.4^{+0.3}_{-0.2}$ \\
        \bottomrule
    \end{tabular}
    \label{tab: assembly}
\end{table}


\begin{table*}
    \centering
    \caption{Comparison of stellar mass evolution of BCGs to observational data. The values are presented as $\log(M_{\star}/M_{\odot})$.}
    \label{tab:bcg_mass_comparison}
    \begin{tabular}{lccc}
        \hline
        \hline
        Redshift & \citet{erfanianfar19}$^a$ & \citet{lin13} & This Work \\
        \hline
        $0.1 < z < 0.3$ & 
        $11.46^{+0.24}_{-0.24}$, $11.61^{+0.2}_{-0.2}$, $11.66^{+0.2}_{-0.2}$ & 
        $-$ & 
        $11.58^{+0.09}_{-0.13}$, $11.83^{+0.09}_{-0.16}$, $12.0^{+0.08}_{-0.09}$ \\
        
        $0.3 < z < 0.65$ & 
        $11.46^{+0.25}_{-0.25}$, $11.57^{+0.27}_{-0.27}$, $11.58^{+0.24}_{-0.24}$ & 
        $-$ & 
        $11.47^{+0.12}_{-0.16}$, $11.65^{+0.13}_{-0.13}$, $11.84^{+0.12}_{-0.12}$ \\

        $0.05$ & 
        $-$ & 
        $11.62$ & 
        $11.62^{+0.23}_{-0.11}$ \\

        $0.3$ & 
        $-$ & 
        $11.56$ & 
        $11.58^{+0.01}_{-0.14}$ \\

        $0.5$ & 
        $-$ & 
        $11.58$ & 
        $11.52^{+0.08}_{-0.16}$ \\

        $0.9$ & 
        $-$ & 
        $11.40$ & 
        $11.41^{+0.08}_{-0.2}$ \\

        $1.1$ & 
        $-$ & 
        $11.28$ & 
        $11.23^{+0.1}_{-0.07}$ \\

        $1.4$ & 
        $-$ & 
        $11.23$ & 
        $11.20^{+0.12}_{-0.14}$ \\

        \hline
         
    \end{tabular}

    \footnotesize $^a$ The three values for Erfanianfar et al. (2019) correspond to BCGs residing in halos with masses of $10^{14.2} \ M_{\odot}$, $10^{14.5} \ M_{\odot}$, and $10^{14.8} \ M_{\odot}$, which are comparable to the Fornax-, Virgo-, and Coma-like divisions used in this work.
\end{table*}



\subsection{Assembly rate history}
\label{sec: ARH}

We calculate the assembly rate history to understand how fast the stellar mass was assembled over the time interval between two consecutive snapshots. This value was calculated by considering the differences in stellar mass of the main progenitor and time between consecutive snapshots. The stellar mass assembly rate can provide insights into how the stellar mass buildup of galaxies has progressed throughout their history, highlighting specific periods of increased or decreased changes in stellar mass content.

\begin{figure*}

	\includegraphics[width=\textwidth]{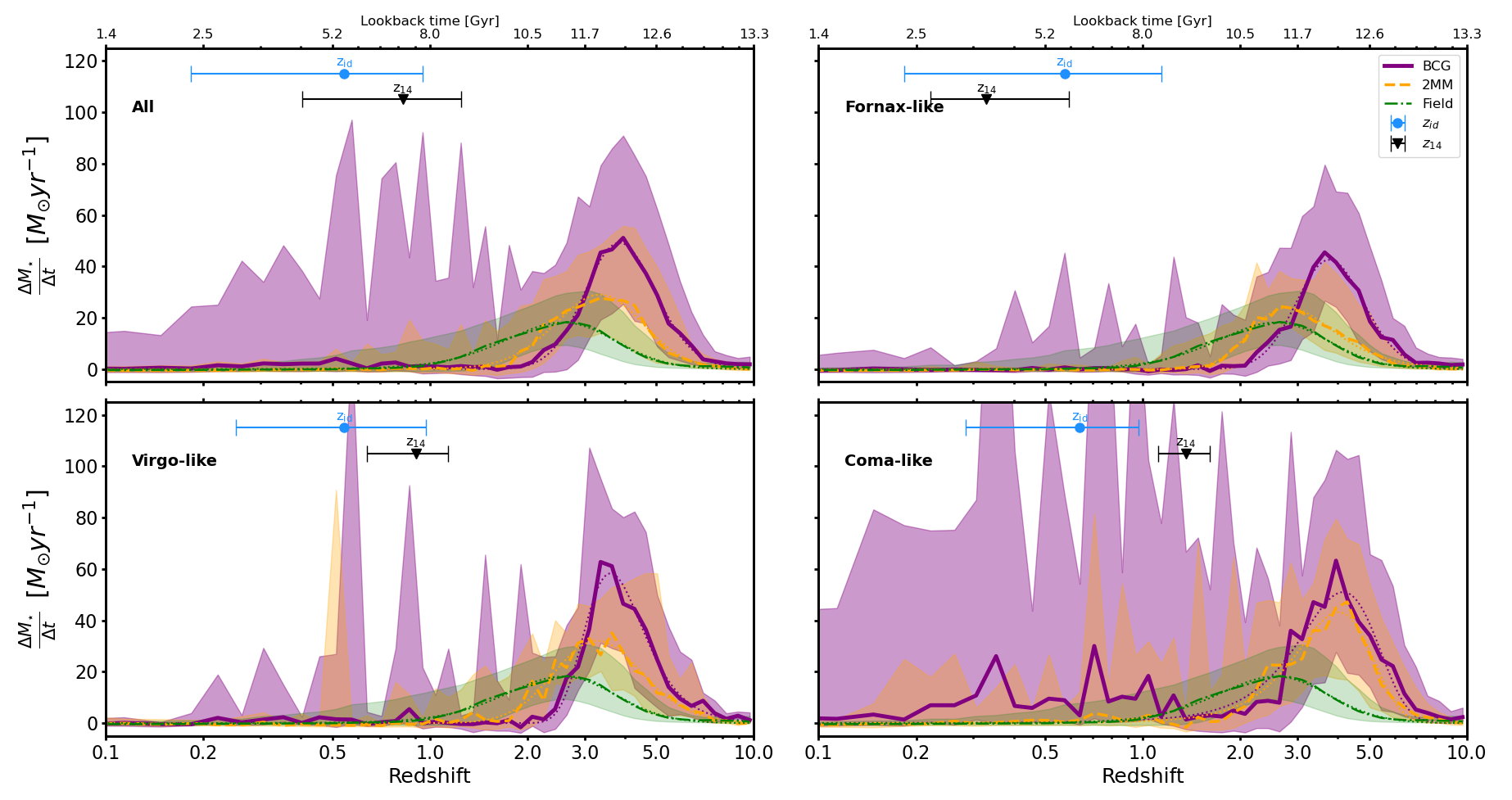}
    \caption{Stellar mass assembly rate histories. The curves and colors follow the same scheme as previous figures. Colored dotted curves are modified Gaussian fit (see Eq. \ref{eq: mod_gauss}). The black inverted triangle with error bars and cyan dot with error bars, indicate the median and inter-quartile range of $z_{14}$, and the median and inter-quartile range of $z_{id}$.    
}
    \label{fig: sm_rate_hist}
\end{figure*}

Figure \ref{fig: sm_rate_hist} presents the stellar mass assembly rate histories. We also added modified Gaussian fits (dotted curves; using the same color scheme) to the data to identify the redshift assembly rate peak redshift and its uncertainty ($z_{peak, AR}^{type}$; see Table \ref{tab: assembly}). The modified Gaussian is given by:

\begin{equation}
\label{eq: mod_gauss}
f(z; A, \mu, \sigma, s) = A \exp\left[-\left(\frac{z - \mu}{\sigma + s \cdot (z - \mu)}\right)^2\right],
\end{equation}

\noindent
where $A$ represents the amplitude, $\mu$ is the mean, $\sigma$ is the standard deviation, and \textit{s} is an additional parameter that introduces asymmetry.

The progenitors of BCGs start to assemble stellar mass earlier, at $\rm 7 \lesssim z \lesssim 8$, and more rapidly than other galaxies, reaching their peak at $z_{peak, AR} \sim 3.8 - 4.2$ for Fornax- and Coma-type clusters, respectively, and then decreasing more steeply than the other galaxy populations up to $\rm z \sim 2.1$. The $\rm 2MMs$ occupy an intermediate position between the BCGs and field galaxies, with those in Coma-like (Fornax-like) structures exhibiting behavior more similar to the BCG (Field galaxy) curve. This behavior is consistent with the downsizing trend: more massive galaxies assemble their stars earlier and over shorter timescales \citep[e.g., ][]{cowie96, neistein06, thomas10}. 

The peaks for BCGs are also on average 40\% higher than $\rm 2MMs$ and field galaxies, despite the fact that the standard deviation is similar (see Table \ref{tab: assembly}). 

The assembly mass rate of the BCGs shows multiple peaks throughout their history. 2MMs show smaller peaks and field galaxies evolve more smoothly. This effect is partly explained by stochasticity. However, the history of the $\rm 2MMs$ is also traced by statistics of the same number of objects, and yet the orange curve representing these objects is less peaky. To test this further, knowing that BCGs are the most massive galaxies at $z=0$, we repeated the analysis normalizing the assembly rate by the final mean mass of each population, and BCGs still showed a more peaky evolution. As for field galaxies, we also repeated the analysis using only 180 random objects, and the result remained smooth, with no peaks. A plausible explanation is that BCGs, by residing at the bottom of the gravitational potential well of their structures, have a higher likelihood of experiencing late-time minor mergers with satellite galaxies that sink toward the center via dynamical friction \citep[e.g.,][]{edwards20, gozaliasl24}. Such events contribute to episodes of rapid stellar mass growth that are not as frequent among other galaxy populations. These features are mainly observed at $\rm z < 2$, approximately 2$\sigma$ after the peak of the assembly mass rate, when the structures begin to reach the mass threshold of our galaxy cluster definition. 

Specifically, the largest of these peaks in the upper percentile occur near $z_{id}$. This is expected, as $z_{id}$ is defined as the redshift at which the last major merger occurs in the main progenitor of the BCG. This behavior varies across structure types: Fornax-like clusters show lower incidence and intensity, while Coma-like clusters display it more frequently and with greater amplitude, often with prominent peaks even in the median curve. This further highlights the complex and distinct evolutionary histories of BCGs compared to other massive galaxies.

\subsection{Formation history}
\label{sec: FH}

Now, we study the star formation histories of the galaxies in our samples, following the same framework used in the previous sections. The star formation history of a galaxy is defined as the ratio between the sum of progenitors' stellar masses at a given snapshot and the final stellar mass of the galaxy under consideration (for definitions, see Section \ref{sec: SMH}). 

\begin{figure*}[ht]

	\includegraphics[width=\textwidth]{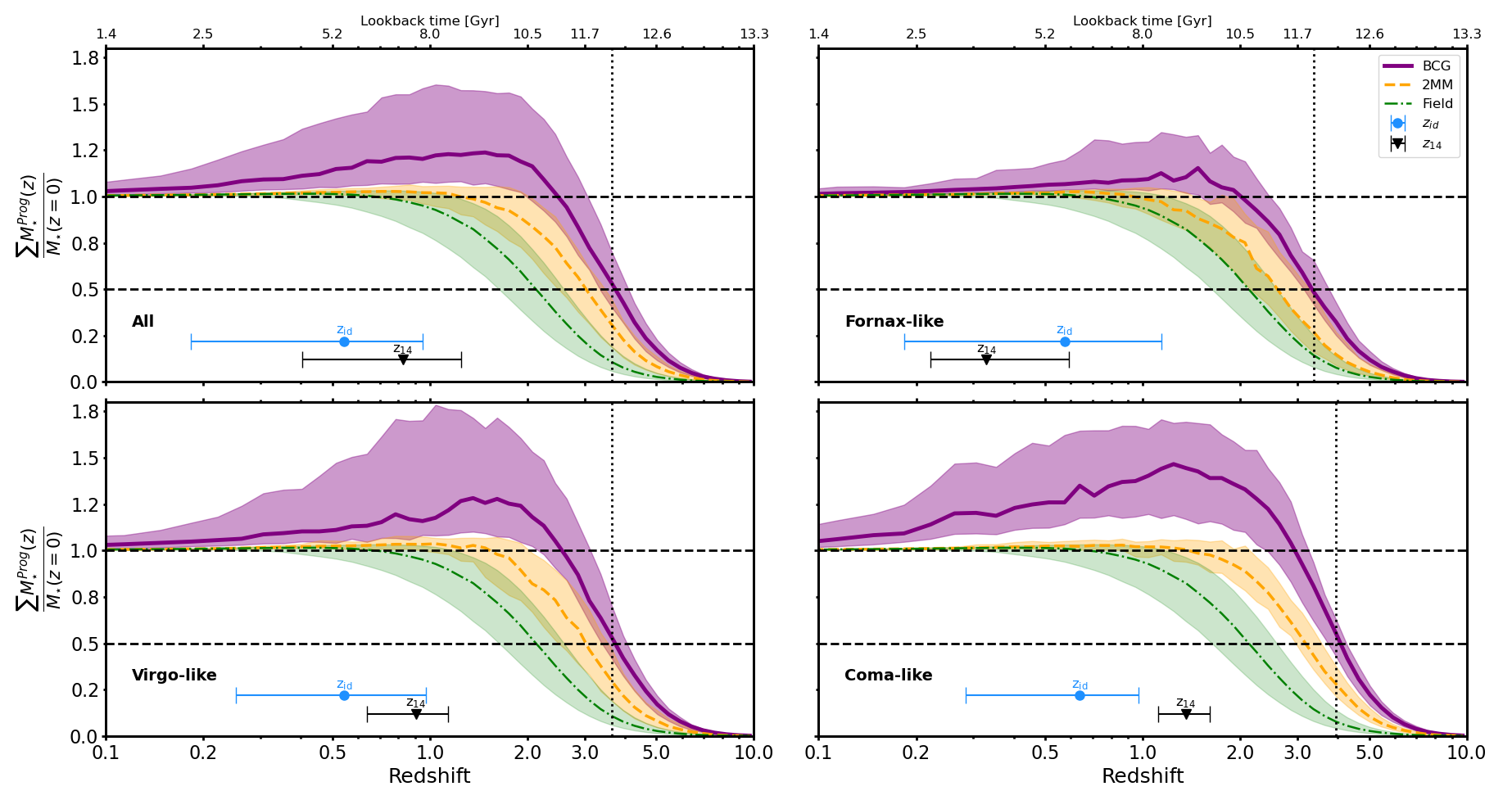}
    \caption{Stellar mass formation histories, i.e., the ratio between the sum of progenitors' stellar masses at a given snapshot and the final stellar mass of the galaxy under consideration. The curves and colors follow the same scheme as in Figure \ref{fig: sm_hist}. The dotted vertical line denotes the median redshift when BCGs formed 50\% of their final stellar mass. The black inverted triangle with error bars and cyan dot with error bars, indicate the median and inter-quartile range of $z_{14}$, and the median and inter-quartile range of $z_{id}$. The curve for BCGs exceeds unity because the total stellar mass of their progenitor galaxies is greater than the stellar mass of the BCG at $z = 0$. This occurs because satellite galaxies approaching the BCG undergo disruption, with their stellar material being deposited into the BCG’s halo. 
}
    \label{fig: sm_form_hist}
\end{figure*}

Figure \ref{fig: sm_form_hist} shows the star formation histories of BCGs, 2MMs, and massive field galaxies. Stars that will end up in BCGs are formed earlier, massive field galaxies later, and 2MMs exhibit an evolutionary pattern that falls between the other two. Table \ref{tab: assembly} shows the redshift at which 50\% of the stars were formed ($\rm z_{50f}^{type}$). Except for Coma-like clusters, the stars in BCGs, 2MM, and Field galaxies, begin to form significantly at $\rm z \lesssim 7, 6, and \ 5$, respectively. The star formation grows steeply, mainly for the BCGs, in the range $\rm z \sim 2.2 - 5$, forming 50\% of their final stellar mass at $\rm z \sim 3.6$ and reaching 100\% by $\rm z = 2.5$. For BCGs inhabiting Coma-like structures, star formation starts even earlier, beginning to rise around $\rm z \sim 8$, reaching 50\% at $\rm z \sim 4$, and 100\% by $\rm z \sim 3$. These results differ somewhat from those found by DLB07, where 50\% of the stars that end up in the BCG are formed by $z \sim 5$ (see their Figure~7). This discrepancy reflects differences in the physical prescriptions adopted in successive versions of \texttt{L-GALAXIES}, particularly in the treatment of satellite galaxies. In DLB07, satellites experienced immediate and complete removal of their hot-gas reservoirs upon entering an FOF group. More likely, the gradual gas removal implemented in later versions of \texttt{L-GALAXIES} leads satellites to retain larger gas (and stellar) masses by the time they merge with the central galaxy. This allows central galaxies to grow more efficiently than in DLB07, where 100\% of the satellites’ hot gas is transferred to the circumgalactic medium immediately and, in cluster environments, AGN feedback prevents this gas from cooling again. We note that differences in other physical prescriptions between the models may also contribute to the remaining discrepancies.

Another interesting characteristic observed is that the sum of the BCG progenitors' masses is greater than the galaxy's final mass over a broad range of redshifts, $\rm z \lesssim 3$ ($\lesssim 2$) for Coma-like (Fornax-like) structures. At lower redshifts, the curves begin to decelerate and decrease smoothly until the ratio stabilizes at unity around $\rm z \sim 0.1$. This can be explained by the fact that the progenitor galaxies of the BCGs undergo significant environmental processes, such as tidal stripping, while this is a less important factor for 2MMs and field galaxies.

In general, the BCG curves start to decline when the structures transition from protocluster to cluster ($\rm z_{14}$), indicating that environmental processes begin to act more strongly on these galaxies that will become part of the BCG. However, for Fornax-like structures, the curve declines earlier, becoming approximately stable by $\rm z \sim 0.9$, before $z_{14}$, suggesting that the most significant stellar mass removal from the progenitors of BCGs in these structures occurred earlier, in the protocluster phase \citep[pre-processing; e.g., ][]{Fujita04, olave-rojas18, ayromlou19, werner22}. For BCGs in Coma-like clusters, the environmental effects are stronger, removing a substantial amount of stellar mass from the progenitors as they move inwards the cluster to be finally cannibalized by the BCG. In these massive structures, the sum of the progenitors' masses rises to $\gtrsim$ 50\% more than the BCG's final mass at $\rm z = 1-2$, as shown in the bottom right panel of Figure \ref{fig: sm_form_hist}.

\begin{figure*}[ht]

	\includegraphics[width=\textwidth]{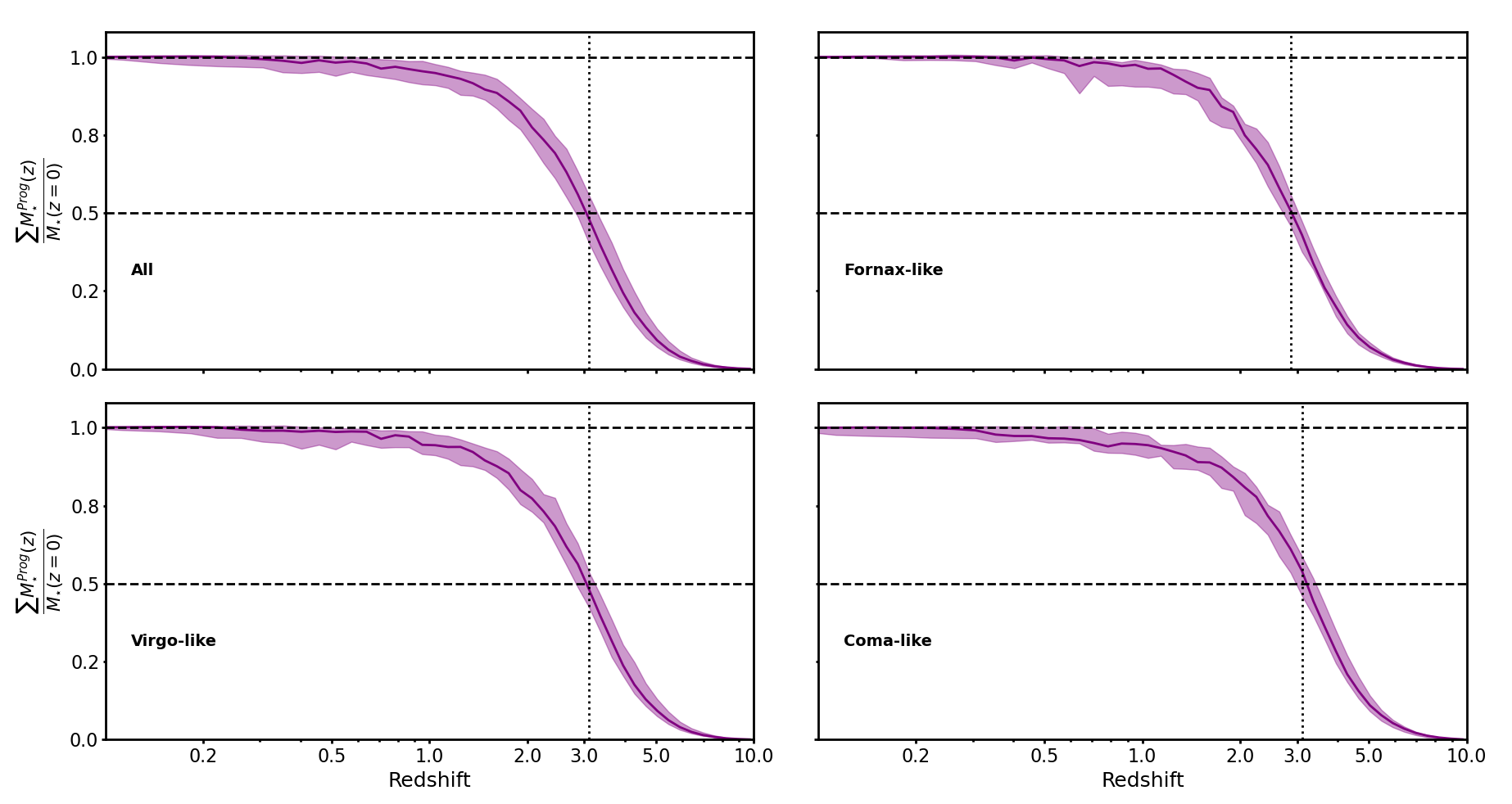}
    \caption{(BCG stellar mass formation histories considering also the halo stellar mass. Differently from Figure \ref{fig: sm_form_hist}, the ratio of the progenitors' mass to the final BCG mass does not exceed unity, indicating that these stars were indeed removed from their host galaxies and remain in the BCG's halo. 
}
    \label{fig: sm_form_hist_plus_halo}
\end{figure*}

To evaluate if this effect is indeed due to the stripping of stellar mass throughout the BCG's history, we repeated this analysis in Figure \ref{fig: sm_form_hist_plus_halo}, now also considering the stellar mass present in the galaxy's halo for the BCGs and their progenitors. In this case, the ratio between the progenitors’ total mass and the final BCG mass no longer exceeds unity. This behavior is a direct consequence of how \texttt{L-GALAXIES} treats stellar stripping: stars tidally removed from satellites are added to the stellar halo of the central galaxy. Figure \ref{fig: sm_form_hist_plus_halo}, therefore, illustrates this built-in process. In A21, stars are added to the stellar halo only when satellites are fully and instantaneously disrupted by tidal forces, so these galaxies do not later merge with the central.

When orphan galaxies reach the pericenter with the central galaxy associated with the main dark matter halo, if the baryonic density within the half-mass radius is lower than the dark matter density of the main halo within the pericenter, the orphan galaxy is tidally disrupted and its stellar component is added to the halo of the central galaxy. The results presented here highlight the importance of this process for BCGs and protoBCGs. Throughout their evolutionary trajectory, a significant number of satellite galaxies spiral towards the cluster center, their dark matter subhalos are accreted by the overwhelmingly more massive main halo, and subsequently, these orphan galaxies are tidally disrupted, with their stellar material being added to the halo of the (proto)BCG, possibly forming the extended stellar envelope and the ICL.


\subsection{Stellar mass origin: Mergers and In Situ star formation histories}
\label{sec: MBISH}

At this point, an interesting question that arises is about the phenomena responsible for star formation throughout these galaxies' histories. In this section, we examine two quantities from the \texttt{L-GALAXIES} model: stellar mass growth from mergers and from secularly \textit{in situ} star formation. We track their evolution using the ratio of each contribution to the total $M_{\star}$ at a given redshift. Although \texttt{L-GALAXIES} also includes starbursts triggered by mergers, they contribute minimally ($\lesssim$ 3\%) and are included together with the stellar mass growth from mergers.

\begin{figure*}

	\includegraphics[width=\textwidth]{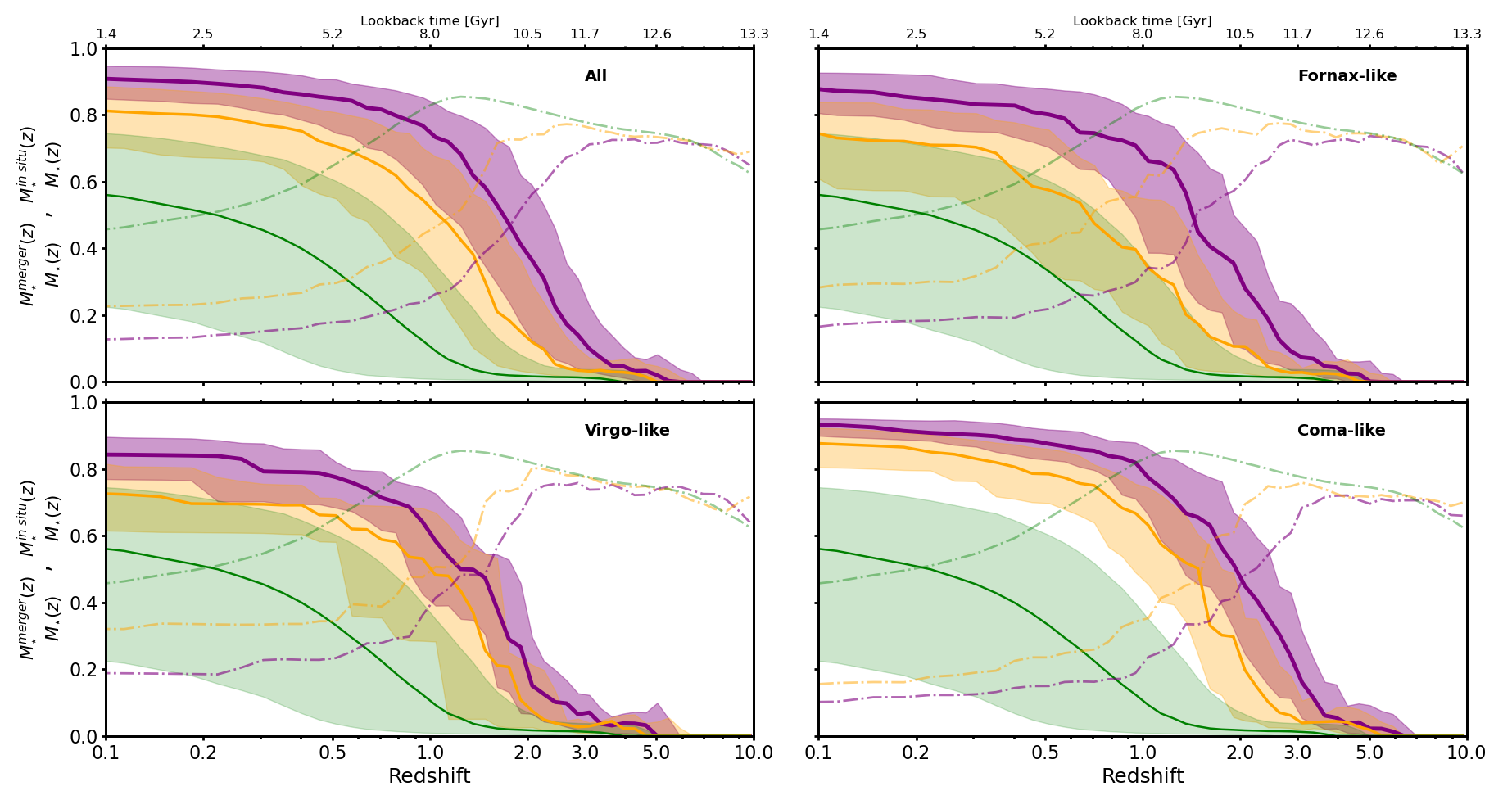}
    \caption{Mergers and \textit{in situ} stellar mass histories. Solid and dot-dashed curves stand for stellar mass fraction assembled due to mergers and \textit{in situ} star formation, respectively, as a function of redshift. Colors follow the same scheme as previous figures. For the sake of visualization, we chose not to include the interquartile range for the \textit{in situ} star formation curves, as they have a similar width as the merger curves.   
}
    \label{fig: sm_origin}
\end{figure*}

Figure~\ref{fig: sm_origin} shows the fractional contribution of stellar mass assembled through mergers (solid lines) and in situ star formation (dot–dashed lines). At $6\lesssim z\lesssim10$, about 80\% of the stellar mass forms in situ from the galaxy’s own cold gas, while mergers become increasingly important at lower redshift. For BCGs the merger contribution rises sharply around $z\!\sim\!3.5$ and dominates below $z\!\sim\!1.8$, earlier than for 2MMs ($z\!\sim\!1.1$) and field galaxies ($z\!\sim\!0.3$). This reinforces the trends in the previous sections: galaxies in denser environments assemble their mass earlier and over shorter timescales. By $z\!\sim\!0.1$, mergers account for $\sim$90\% of the BCG stellar mass, compared with $\sim$80\% for 2MMs and $\sim$55\% for field galaxies.

Despite the systematic difference between BCGs and other massive galaxies, their curves show similar shapes. The curves stabilize after most host clusters exit the protocluster phase, when BCGs—already the dominant central galaxy (see Section~\ref{sec: dominance})—continue to assemble stellar mass mainly through minor mergers, contributing $\sim$5\% of their $z=0$ mass. Residual growth is also seen for the 2MMs, which occur through a combination of processes: (i) direct mergers between satellite galaxies whose dark-matter subhaloes merge in the underlying N-body simulation, and (ii) mergers within smaller subhaloes in which the 2MM was the central galaxy prior to being incorporated into the main cluster halo. Both channels allow merger-driven mass build-up before the system becomes fully dynamically relaxed.

For field galaxies, mergers contribute significantly only at $z\lesssim 1$. These galaxies inhabit lower mass FOF groups at $z = 0$ (see Section~\ref{sec: defs}), and their halos were even less massive at higher redshifts, making them more isolated and less prone to mergers. 
At later times, however, some occupy halos where group environments begin to form and mergers become increasingly relevant.

There are significant differences in the evolution of stellar mass growth origin for galaxies residing in Fornax-, Virgo-, or Coma-like structures. For BCGs, the contribution from mergers in Fornax-like structures begins to grow more strongly later, around $\rm z \sim 2.8$, and stabilizes at $\rm z \sim 0.6$. In contrast, for more massive structures like Coma-like clusters, this contribution increases sharply as early as $\rm z \sim 3.8$ and stabilizes at $\rm z \sim 1$. Furthermore, the discrepancy between massive cluster galaxies and field galaxies is greater. Coma-like clusters form earlier, as indicated by $\rm z_{14}$ in Table \ref{tab: assembly}. These structures exhibit a higher galaxy density compared to less massive halos, even at high redshifts, creating a favorable environment for mergers. This process is more pronounced and occurs earlier for (proto)BCGs regardless of the structure's mass, as shown in Figure \ref{fig: sm_origin}. This suggests that these galaxies already reside in a denser region within the protocluster structure at redshifts as high as $\rm z \sim 4$ (see Section \ref{sec: density_ratio}).

\subsection{Bulge stellar mass histories}
\label{sec: DBH}



Another interesting question is how stars are distributed between the disk and bulge, as the mass fraction in each component is directly linked to morphology and indirectly to the galaxy’s star formation rate. Bulge-dominated galaxies are typically elliptical, while disk-dominated systems are spiral or lenticular; in the local Universe, ellipticals tend to be redder and quiescent or quenching, whereas spirals are bluer and actively forming stars \citep[e.g.,][]{kennicutt98, blanton09, tachella19}. 

\begin{figure*}

	\includegraphics[width=\textwidth]{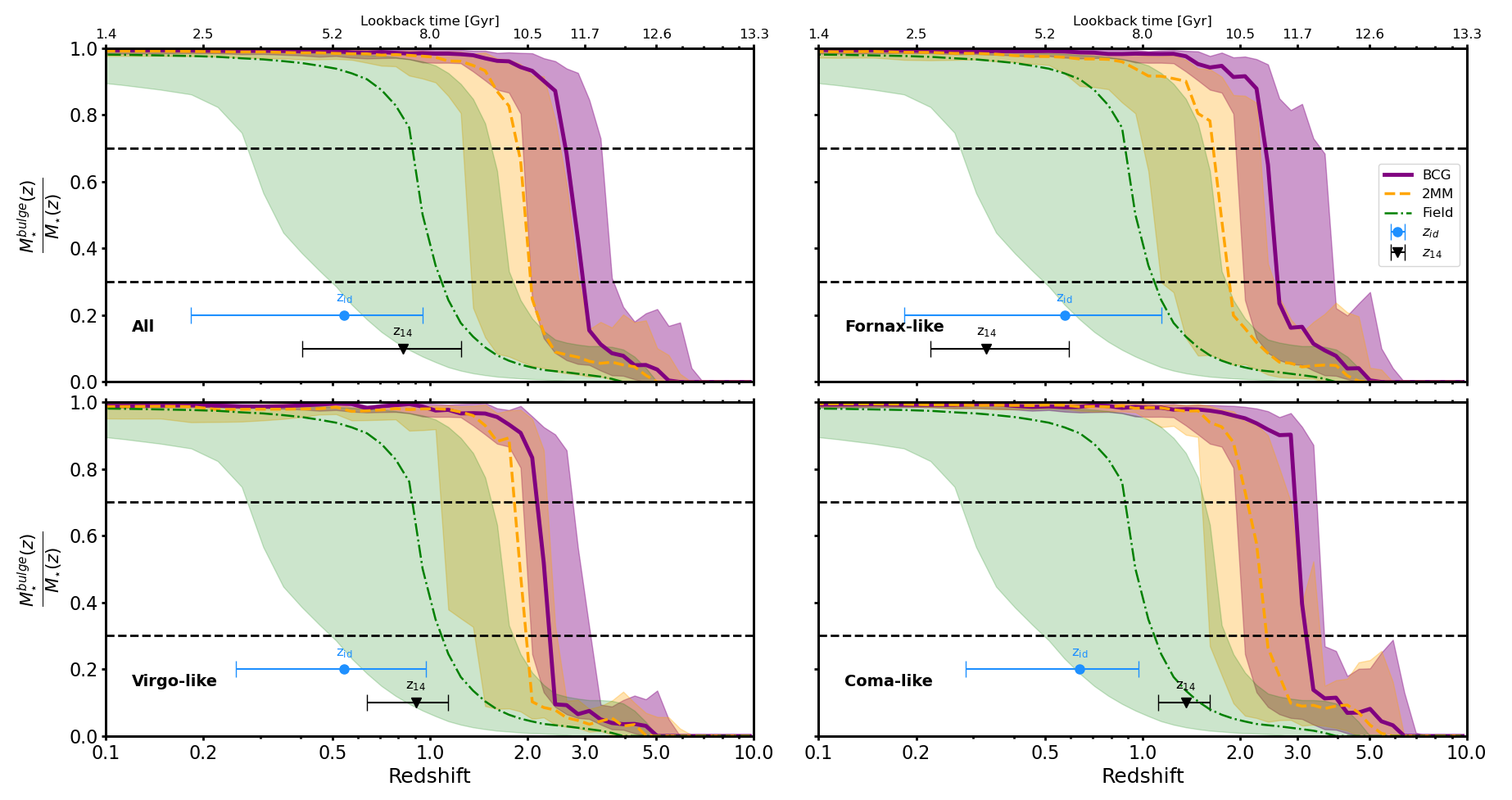}
    \caption{Bulge stellar mass content histories. The curves and colors follow the same scheme as previous figures. The black inverted triangle with error bars and cyan dot with error bars, indicate the median and inter-quartile range of the cluster formation redshift ($z_{14}$), and the median and inter-quartile range of the identity redshift ($z_{id}$).   
}
    \label{fig: sm_bulge}
\end{figure*}

Figure \ref{fig: sm_bulge} shows the history of the bulge stellar mass fraction relative to the total as a function of redshift for BCGs, 2MMs, and field galaxies. There is a distinction between the build-up of the bulge stellar mass between the different types of massive galaxies--However, at low-z ($\rm z < 0.4$), massive galaxies are bulge-dominated, regardless their environment. For all three curves, there is a point where the increase in the bulge stellar fraction becomes very steep, highlighting that the morphological transition from a disk-type to an elliptical galaxy occurs rapidly in these massive galaxies. This sharp transition reflects the internal treatment of disk instabilities and merger events in \texttt{L-GALAXIES}, which transfer stellar and gaseous material to the bulge component once specific stability thresholds are exceeded. Field galaxies transit more gradual--although not entirely smooth--and later than cluster galaxies. This drastic increase happens at $\rm z \sim 3$ for BCGs, although it is smaller for Fornax-like clusters. For 2MM and field galaxies, this transition occurs later across all cases, generally around $\rm z \sim 2.1 \ and \ 1.1$, respectively. Comparing this with the merger history (Figure \ref{fig: sm_origin}), we observe a clear correspondence between the transition to bulge-dominated galaxies and the epoch when mergers become the dominant contributor to stellar mass assembly. This connection is expected, as bulge growth in galaxies is closely linked to merging processes, discussed above. Interestingly, this period also coincides with the post-peak phase of the stellar mass assembly rate (Figure \ref{fig: sm_rate_hist}), indicating that the bulge component of these galaxies grows significantly after they have gone through the most intense phase of stellar mass build-up.

Considering the median values and the inter-quartile ranges of $z_{14}$ and $z_{id}$, we conclude that the protoBCGs (also for the other massive galaxies) transform into elliptical galaxies a considerable time before the structure reaches the cluster mass threshold of $\rm 1.5 \times 10^{14} \ M_{\odot}$, and even more time before when the BCGs reach their identity, as defined in Section \ref{sec: SMH}.

\section{BCG dominance} \label{sec: dominance}

At $z \sim 0$, BCGs are the most massive and luminous galaxies in their clusters, usually located near the minimum of the cluster’s gravitational potential well. This section aims to understand how BCGs evolve to become the dominant galaxies in their structures. 

We analyze their evolutionary paths relative to the most massive galaxy in the same structure at each redshift, excluding the BCG progenitor itself. We denote this galaxy as $\mathrm{MM_z}$ (``most massive at redshift $z$''). Operationally, we first identify all galaxies that belong to each cluster at $z = 0$ based on their FOF membership, then trace all their progenitors back in time using the merger trees. At each snapshot, $\mathrm{MM_z}$ corresponds to the most massive progenitor that is not part of the BCG’s own merger tree. This ensures that $\mathrm{MM_z}$ always represents a galaxy that will end up in the same $z=0$ cluster, rather than a massive galaxy in the surrounding field.

In the following, we analyze the evolution of the stellar mass ratio and distance between (proto)BCGs and $\rm MM_{z}$ (Section \ref{sec: sm_ratio}), the physical offset between (proto)BCGs and the minimum of the potential well (Section \ref{sec: offset}), and the ratio of the density of galaxies and stellar mass between (proto)BCGs and $\rm MM_z$ (Section \ref{sec: density_ratio}). We use either comoving or proper distances depending on convenience, in order to facilitate comparisons with results from other works, as indicated in the subsections below.

\subsection{(Proto)BCG and $\rm MM_{z}$ Stellar Mass Ratio}
\label{sec: sm_ratio}

\begin{figure*}

	\includegraphics[width=\textwidth]{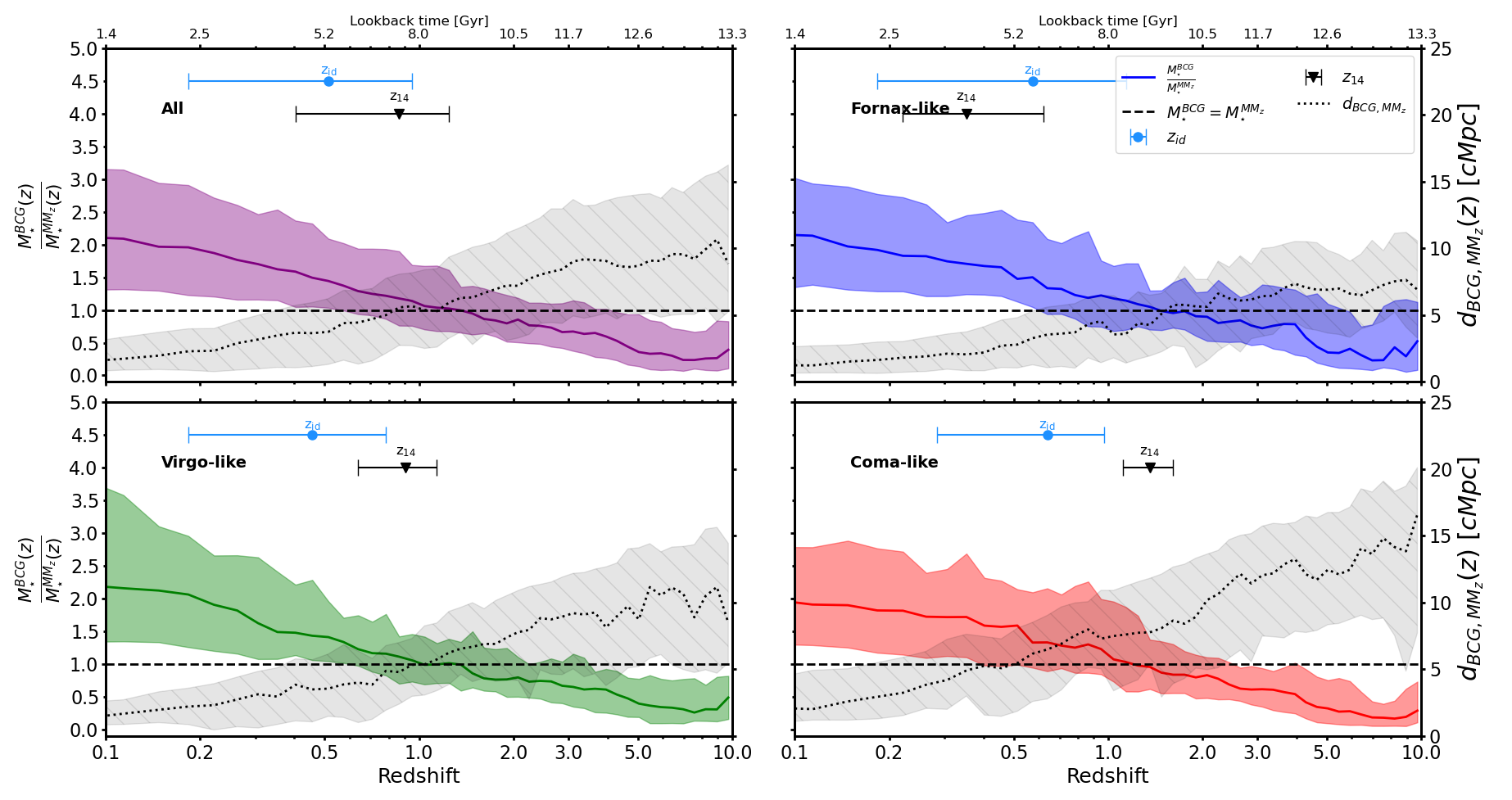}
    \caption{Left y-axis: the stellar mass ratio between the BCG and the most massive galaxy at a given redshift ($MM_{z}$) as a function of redshift. Right y-axis: the comoving distance between the BCG and $MM_{z}$ ($d_{BCG, \ MM_{z}}$). The four panels are for all structures in our sample (top-left), only Fornax-like (top-right), Virgo-like (bottom-left), and Coma-like (bottom-right), as defined in Section \ref{sec: defs}. The solid purple, blue, green, and red curves give the median of the mass ratio (left y-axis). The dashed horizontal line marks a mass ratio of unity. The dotted black curve denotes the comoving distance between the BCG and $MM_{z}$. The shaded areas encompass the 25th and 75th percentiles. The black inverted triangle with error bars and cyan dot with error bars, indicate the median and inter-quartile range of the cluster formation redshift ($z_{14}$), and the median and inter-quartile range of the identity redshift ($z_{id}$), respectively.    
}
    \label{fig: sm_ratio}
\end{figure*}

Figure \ref{fig: sm_ratio} explores the mass ratio between the BCGs and the $\rm MM_{z}$ (left y-axis), and the comoving distance between these two galaxies ($\rm d_{BCG, \ MM_{z}}$, right y-axis), as a function of redshift. We adopt comoving units for the distance in order to directly compare our results with the effective radii of protoclusters reported by \citet{chiang13}.

Generally, the main progenitors of the BCGs in the protocluster phase ($\rm z > z_{14}$), i.e., protoBCGs, are not the most massive galaxies in the structure, as $\rm M_{\star}^{BCG}/M_{\star}^{MM_{z}} <$ 1. When protoclusters begin to reach cluster mass, according to our definition, at $z_{14}$, the BCGs become the most massive galaxy. This transition occurs typically at $\rm z \sim 1.3$ independent of the cluster's mass. This is consistent with the findings of \citet{dalal21}, who analyzed an observed sample of BCGs at $\rm z < 1$ and demonstrated that BCGs are `special' galaxies in the sense that they follow a distinct mass distribution, independent of other massive cluster galaxies. In particular, they report an average mass gap between BCGs and the second-ranked galaxy of $\rm \langle \log(M_{BCG}/M_{MM_{z}}) \rangle \sim 0.18$ for $\rm z < 1$\footnote{In their observational analysis, the second most massive galaxy is defined at the observed redshift, corresponding to the most massive galaxy in the structure after excluding the BCG at that epoch. This is therefore equivalent to our definition of $\rm MM_{z}$.}. Here, we find a comparable mass gap of $\rm \langle \log(M_{BCG}/M_{MM_z}) \rangle = 0.27 \pm 0.02$ at $\rm z < 0.65$, and $\rm 0.25 \pm 0.02$ for $0.65 < z < 1$. Although our values are slightly higher, both the observational and simulated results consistently indicate that BCGs exhibit a systematically larger stellar mass than the $\rm MM_{z}$ at $\rm z < 1$. Figure \ref{fig: sm_ratio} also shows that, by $z_{id}$, BCGs already have about 50\% more mass than the $\rm MM_{z}$, and by $\rm z \sim 0$, they have more than double the stellar mass of the second most massive galaxy.

The black dotted curves show that during the protocluster phase, $\rm d_{BCG, \ MM_{z}}$ is typically large--often exceeding 5 cMpc. For Coma-like clusters, this distance is even greater, reaching approximately 13 cMpc at $\rm z \sim 5$, while for Fornax- and Virgo-like clusters, $\rm d_{BCG, \ MM_{z}} \sim 7 \ and \ 8 \ cMpc$, respectively, at the same redshift. This behavior is understandable, given that Coma-like protoclusters are larger structures, with an effective radius of around 10 cMpc at $\rm z = 5$, as found by \citet{chiang13} using data from the Millennium simulation. In Figure \ref{fig: structures}, we show the spatial distribution of galaxy members for two structures across three different redshift bins, highlighting the positions of the (proto)BCG and $\rm MM_{z}$. In these examples, we note that at high redshifts the two galaxies are more widely separated, often residing in distinct clumps within the protocluster.

These results indicate that, at high redshifts, selecting the BCG progenitor in protoclusters should not rely solely on a mass criterion, but rather on a combination of mass and local density, as we will show in Sections \ref{sec: offset} and \ref{sec: density_ratio}.

\subsection{Offset of (proto)BCGs from the bottom of the potential well} \label{sec: offset}


To calculate the gravitational potential field of our structures, we considered the baryonic matter, which includes the stellar mass, black hole mass, cold and hot gas mass, and stellar mass in the halo of each galaxy member of the structure. Additionally, we included all dark matter subhalos that comprise the (proto)cluster. The mass of each dark matter subhalo was calculated by multiplying the number of dark matter particles by the particle mass (see Section \ref{sec:data}), as suggested by \citet{yates17}\footnote{Calculating the subhalo mass using the number of dark matter particles is more robust than directly using the $\rm M_{200}$ value provided by L-GALAXIES in specific situations, such as fly-bys, due to the intrinsic assumption of symmetry in the latter's calculation.}. Using this information, the gravitational potential field was computed for each structure throughout its history on a 3D grid, divided into a sufficient number of voxels to ensure that the maximum positional uncertainty — set by the voxel size, which defines the furthest distance between the voxel center and the true minimum within that voxel — remains below 10 ckpc. We opt to compute physical offsets in order to directly compare with observational results further in this section.

\begin{figure*}

	\includegraphics[width=\textwidth]{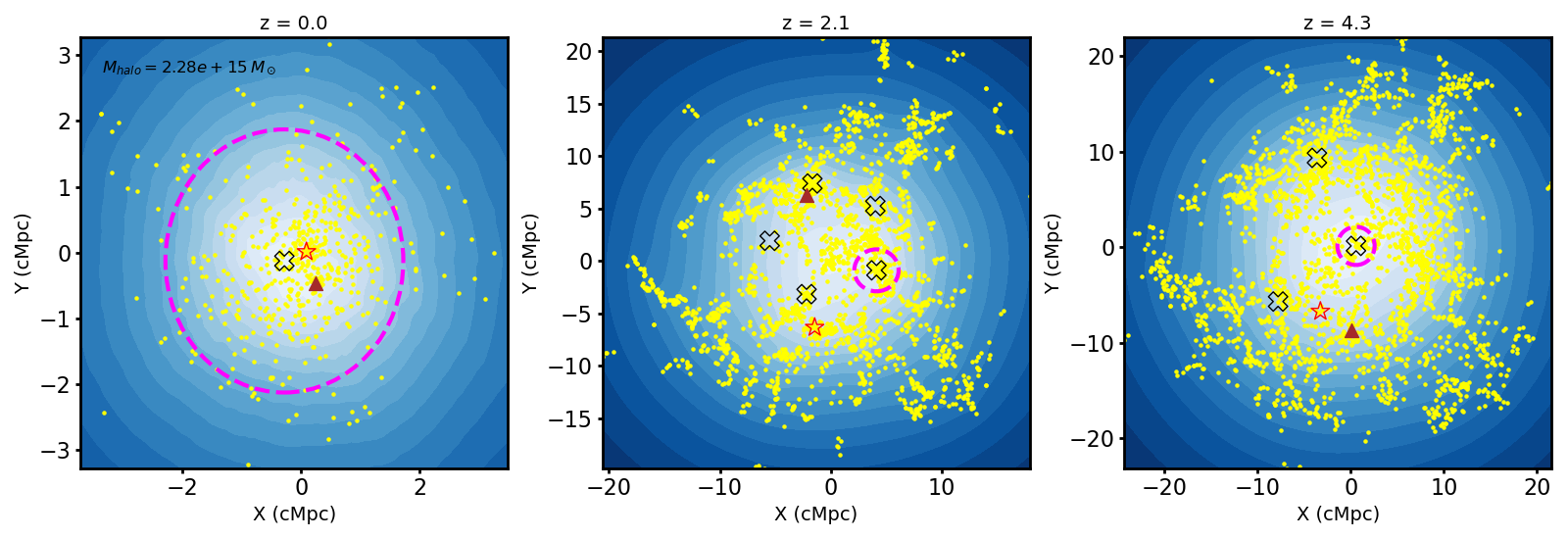}
	\includegraphics[width=\textwidth]{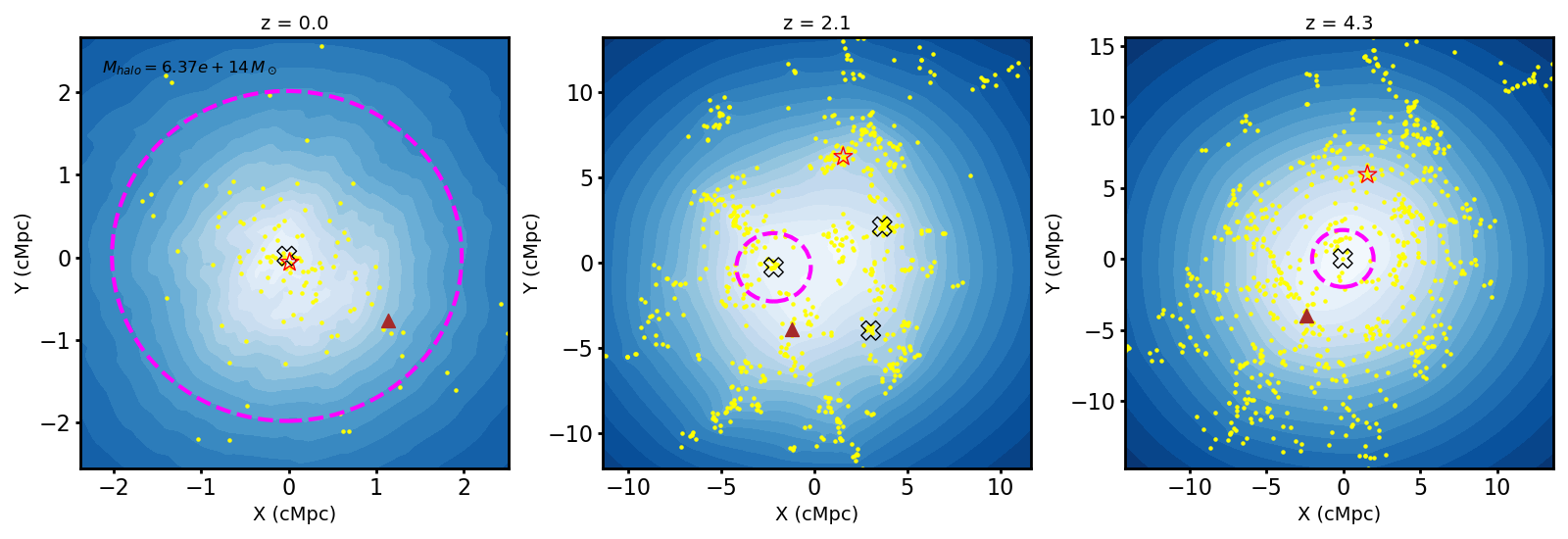}
	\includegraphics[width=\textwidth]{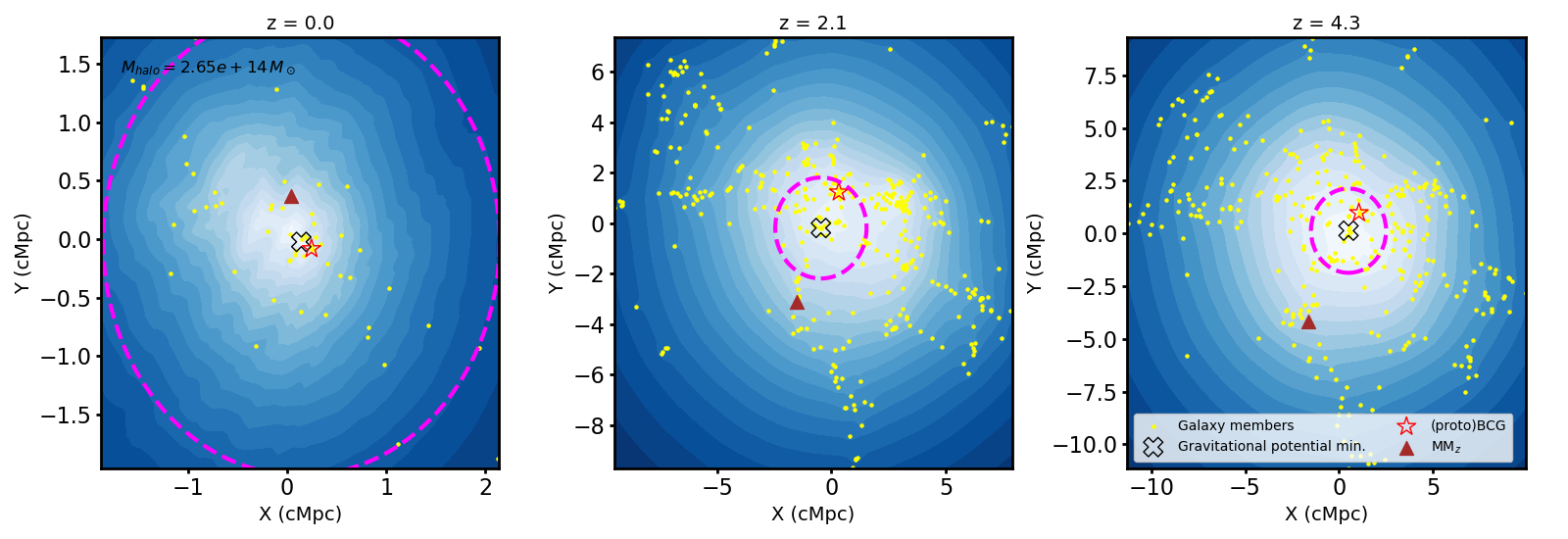}
    
    \caption{Each row shows the same structure at different evolutionary stages, specifically at redshifts $z = 0$, $2.1$, and $4.3$, as indicated at the top of each panel. First, second, and third rows show one example of a Fornax-, Virgo-, and Coma-like structure, respectively. The $\rm M_{halo}$ at $z = 0$ is shown in the top-left corner of the leftmost panel. The colormap represents the gravitational potential field, where darker blue corresponds to shallower (higher) potential values and white indicates deeper (lower) regions. The (proto)BCG is marked by an open red star, while $\rm MM_{z}$ is shown as a brown triangle, and other member galaxies as yellow circles. The global and local minima of the gravitational potential field are indicated by black `X' markers. The global minimum is located at the center of the dashed magenta circle with a radius of 2 cMpc. Local minima are identified within 1 pMpc--corresponding to 1.0, 3.1, and 5.3 cMpc in comoving units for the examples shown in the figure--from the global minimum and with potential values at least 90\% as deep as the global minimum.
}
    \label{fig: structures}
\end{figure*}

At the protocluster stage, these structures are more diffuse, distributed across multiple density clumps that will eventually merge later in their history \citep{overzier16}. With this in mind, in addition to identifying the global minimum of the gravitational potential, we also identify local minima located at least 1 pMpc away from the global or from any other previously detected local well, and whose potential depth is at least 90\% of the global minimum. Figure \ref{fig: structures} shows three examples of the spatial distribution of structures at different evolutionary stages, highlighting the gravitational potential field with its global and local minima, the distribution of member galaxies, the most massive galaxy at each redshift ($\rm MM_{z}$), and the (proto)BCGs. The figure shows that, galaxies in a protocluster are spread over tens of comoving megaparsecs across different gravitationally bound clumps that will eventually concentrate mainly within a region of radius $\rm \sim 1 \ cMpc$.

\begin{figure*}

	\includegraphics[width=\textwidth]{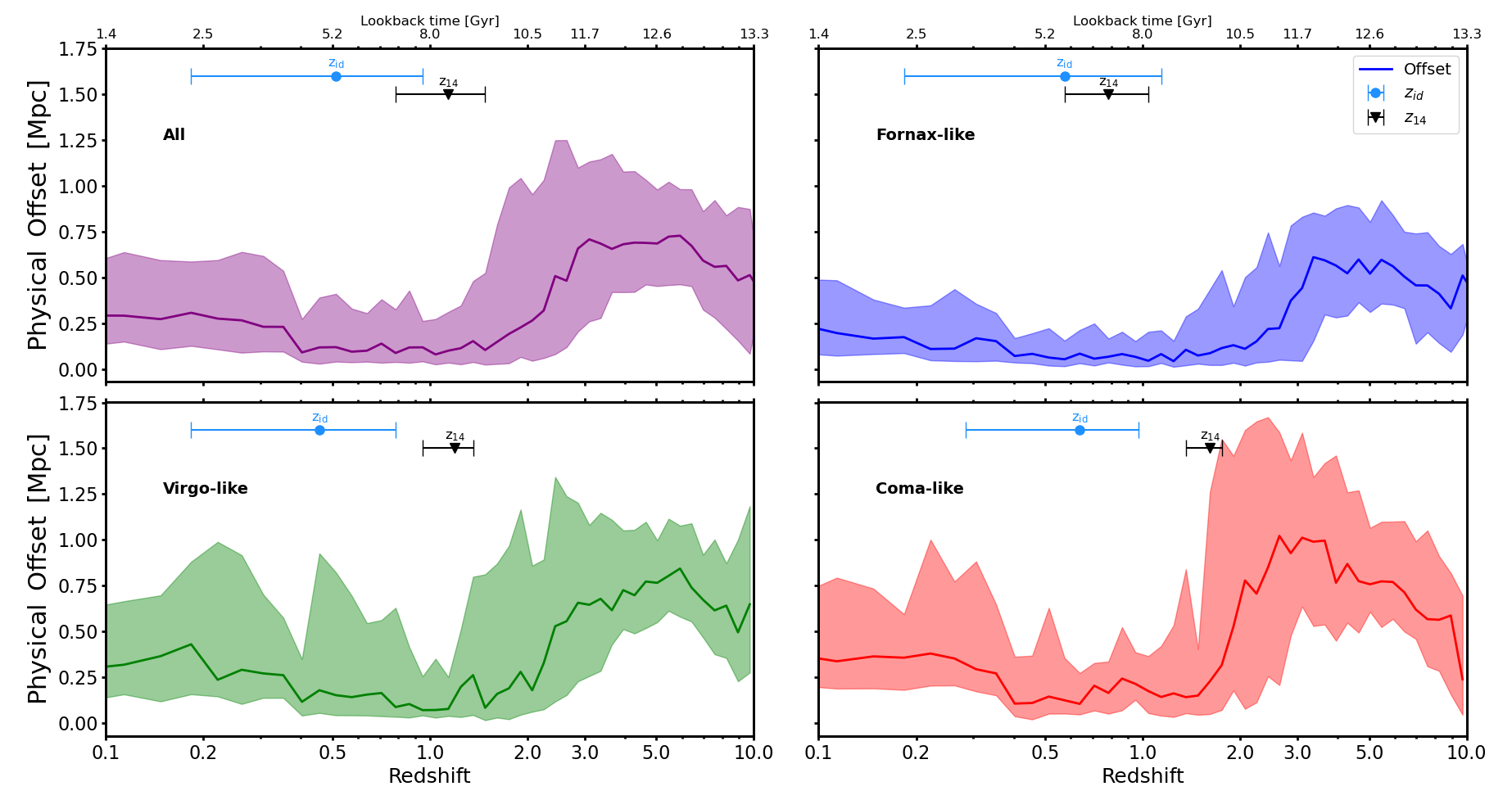}
    \caption{Median physical offset (pMpc) of the (proto)BCG and the potential well's minima as a function of redshift. As previously, plots are divided in four panels: all structures in our sample (top-left), only Fornax-like (top-right), Virgo-like (bottom-left), and Coma-like (bottom-right), as defined in Section \ref{sec: defs}. The shaded areas encompass the 25th and 75th percentiles. The black inverted triangle with error bars and cyan dot with error bars, indicate the median and inter-quartile range of the cluster formation redshift ($z_{14}$), and the median and inter-quartile range of the identity redshift ($z_{id}$). 
}
    \label{fig: offset}
\end{figure*}

Figure \ref{fig: offset} shows the 3D physical offset between the BCG and the closest potential well minimum as a function of redshift. Considering all the structures analyzed (top-left plot), the offset between the (proto)BCG and any minimum of the gravitational potential field evolves in three distinct stages. For $\rm z > 3$, the offset is approximately $\sim$ 0.75 pMpc; in the range $\rm 1.5 \lesssim z \lesssim 3$, it decreases to around $\sim$ 0.1 pMpc, remaining roughly constant until $\rm z \sim 0.4$; and finally, it slightly increases again to $\sim 0.25$ pMpc up to $\rm z = 0$. These stages are observed to a greater or lesser extent depending on the final mass of the cluster, as seen in Fornax-like (upper right plot) and Coma-like (lower right plot) structures, respectively. Note that the offset stabilizes at its lowest value, at around z = $z_{14}$, when the structures start to reach the transitional mass threshold between galaxy (proto)clusters and mature clusters. Part of this trend naturally arises from the definition of central galaxies in \texttt{L-GALAXIES}, where the type~0 galaxy is assumed to reside at the center of the main dark matter subhalo. The quantitative offsets therefore reflect both the model implementation and the underlying physical expectation of dynamical relaxation.

To fairly compare our results with observations, we also computed the projected (2D) offset between (proto)BCGs and the minimum of the gravitational potential, which corresponds to scaling the 3D offset by a factor of $\sqrt{2/3} \approx 0.82$. This reduces the offset by approximately 0.1 pMpc, with typical values ranging from 0.05 to 0.4 pMpc. This range is consistent with offsets reported between BCGs and X-ray peaks in observed clusters \citep[e.g.,][]{oguri18, chu21, depropris21}, which generally lie below 0.1 pMpc, with some cases reaching 0.3–0.5 pMpc. Similarly, \citet{ding25} studied the projected offset between 189 BCGs and cluster centers defined by the Sunyaev-Zeldovich (SZ) effect, used as a proxy for the dark matter distribution, within $\rm 0.1 < z < 1.4$. BCGs were identified using the CAMIRA cluster finder \citep{oguri14, oguri18} in the HSC-SSP photometric survey \citep{aihara18}, matched to the ACT SZ cluster catalog \citep{hilton21}. They found that the offset distribution is well described by two components: one well-centered ($\sim$ 75\% of the sample), peaking at $\sim 0.15$ pMpc with a tail up to 0.5 pMpc, and another due to interacting structures or a minor contribution due to observational artifacts ($\sim$ 25\% of the sample), peaking at 0.385 with a tail up to $\sim$ 1 pMpc. These results are in good agreement with those presented here.

One caveat of this analysis is that we use the subhalo distribution (see Section \ref{sec: merger_trees}) rather than the individual dark matter particles. This introduces uncertainties in the determination of the gravitational potential field, which ultimately affect the offset measurement.

\subsection{Ratio of galaxy and stellar mass density around (proto)BCG and $\rm MM_{z}$} \label{sec: density_ratio}

As discussed in Section \ref{sec: sm_ratio}, (proto)BCGs only become the most massive galaxy within the structure at around $\rm z \sim 1.3$, when these structures start reaching the mass threshold of $1.5 \times 10^{14} \ M_{\odot}$ to be classified as galaxy clusters according to the definition we have used in this paper. Given this context, a key question arises: What drives protoBCGs to increase their mass to become the most massive galaxy, reaching 1.5 to 3.5 times the mass of the second most massive galaxy by $\rm z \sim 0$?

One way to address this question is by examining the environment in which these galaxies are situated. ProtoBCGs may significantly increase their mass compared to other galaxies throughout their history, because they reside in regions with higher concentrations of matter, a situation closely linked to the proximity of protoBCGs to the bottom of the gravitational potential well (Section \ref{sec: offset}). This advantageous position enables protoBCGs to grow by accreting smaller galaxies in their vicinity. The merger process, in addition to adding to the mass of the accreted galaxy, also brings in gas, fueling starbursts episodes and in-situ star formation within the protoBCGs, particularly at higher redshifts, when wet mergers were more common.

\begin{figure*}

	\includegraphics[width=\textwidth]{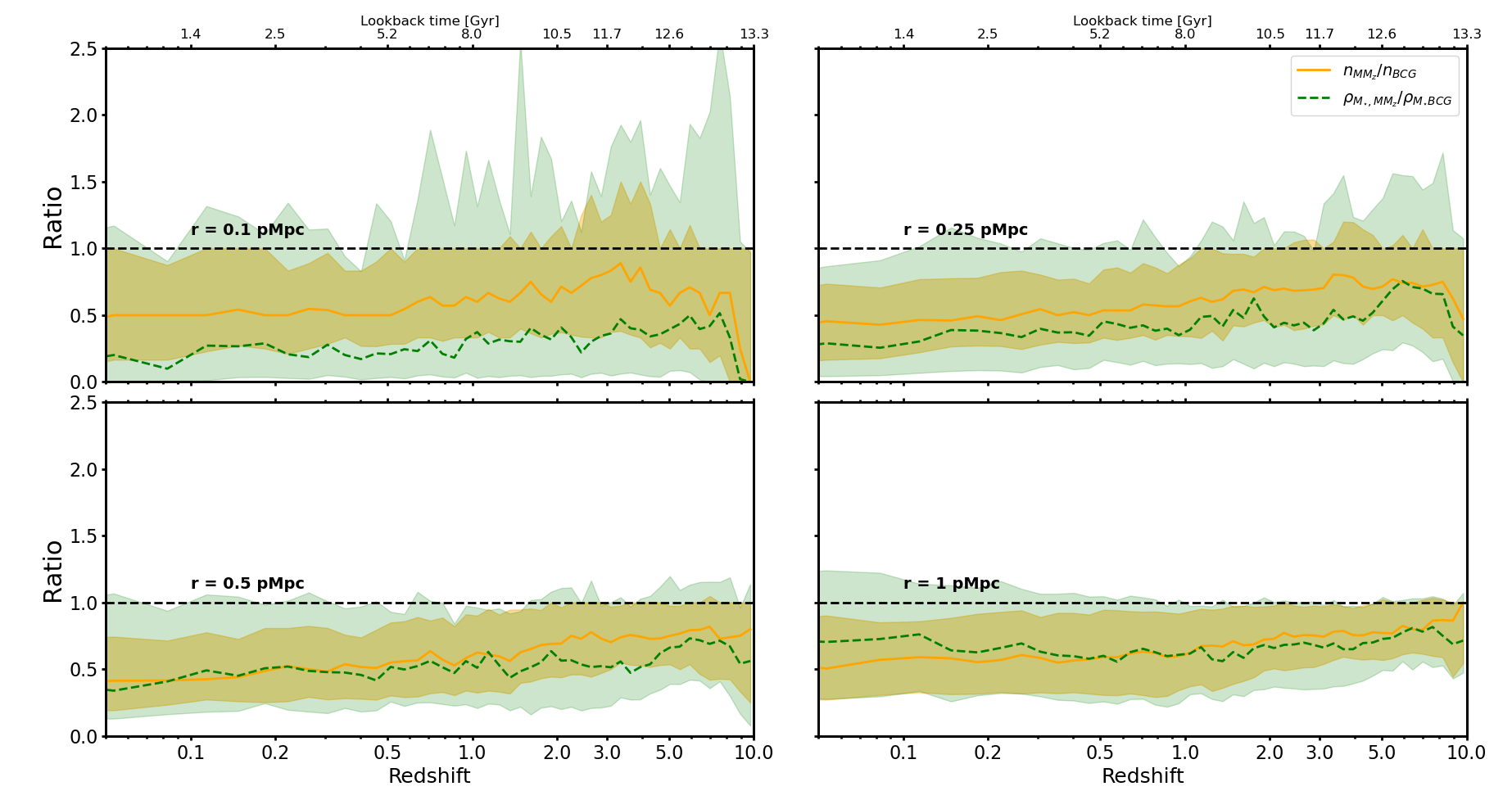}
    \caption{Median galaxy density ratio (solid orange line) and stellar mass density ratio (dashed green line) between the most massive galaxy in the structure at a given redshift (excluding the (proto)BCGs) and the (proto)BCG as a function of redshift. Each plot corresponds to densities calculated within a sphere of different radii, as indicated in each plot. The density calculation excludes the galaxy at the center, i.e., the (proto)BCG or $\rm MM_{z}$ itself. The black dashed horizontal line indicates a ratio equal to unity. The shaded areas represent the range between the 25th and 75th percentiles, while the lines indicate the median.  
}
    \label{fig: density_ratio}
\end{figure*}

Figure \ref{fig: density_ratio} presents the ratio of the galaxy number density (solid orange curves) and galaxy stellar mass density (dashed green curves) centered on the $\rm MM_{z}$ and the (proto)BCG, within spheres of four different radii — 0.1, 0.25, 0.5, and 1 pMpc. When calculating the density, we do not include the contribution of the galaxy at the center of the sphere from which the measurement is made, i.e., the protoBCG or the $\rm MM_{z}$ itself. The figure shows that (proto)BCGs consistently occupy regions with higher galaxy and stellar mass densities than do $\rm MM_{z}$ throughout cosmic history. The disparity in local density increases over time, with galaxy number density ratio doubling between $\rm z = 5$ to $\rm z \sim 0$ and reaching up to five times more in terms of stellar mass density for a smoothing of $\rm r = 0.1 \ pMpc$. This figure shows the results for all clusters; we also performed this analysis by separating the structures into Fornax-, Virgo-, and Coma-like, and found the same behavior in each mass bin. The merger rate depends both on mass density and the local velocity dispersion. Comparing Figure \ref{fig: density_ratio} and Figure \ref{fig: sm_origin} indicates that the merger rate is not slowed down much by the higher velocity dispersion.

Thus, (proto)BCGs indeed reside in environments conducive to rapid mass growth throughout their history, even compared to other galaxies that were more massive than them in the past. This, therefore, is the driver that makes (proto)BCGs the dominant galaxies in their structures by $\rm z \sim 1$.

\section{Summary and Conclusions} \label{sec:conclusions}

In this work, we analyze the stellar mass assembly history of dominant galaxies (i.e., BCGs) in 180 galaxy clusters at $\rm z = 0$, leveraging the \texttt{L-GALAXIES} semi-analytic models of galaxy formation and evolution \citep{ayromlou21}, applied to the dark matter-only Millennium simulation. The analyses are conducted in a comparative manner, considering other massive member galaxies within the structures as well as massive field galaxies. To understand the impact of the host dark matter halo, we carry out our analysis in bins of halo mass of the structure at $z = 0$ in which these galaxies reside.

We define galaxy clusters as structures with halo mass $\rm M_{DM} \geq 1.5 \times 10^{14} \ M_{\odot}$. Structures with halo mass below this threshold but that will exceed it at some future point in the simulation are defined as protoclusters, and the main progenitor of the BCG at this stage is referred to as a protoBCG.

BCGs assemble their mass slower than do other massive galaxies, reaching 50\% of their final stellar mass later, around $\rm z \sim 0.7$, when most galaxy clusters have already evolved past the protocluster phase (Figure \ref{fig: sm_hist}). However, (proto)BCGs are more massive throughout their entire history than the second most massive galaxy defined in the structures at $\rm z = 0$ ($\rm 2MM$) and other massive field galaxies (Figure \ref{fig: sm_abs_assembly}). This difference becomes more pronounced for the most massive Coma-like clusters.

The stellar mass assembly rate for BCGs peaks around $\rm z \sim 4$, well before the structure has reached the galaxy cluster mass threshold of $\rm 1.5 \times 10^{14} \ M_{\odot}$, while other massive galaxies in clusters and field reach maximum rate at later times (Figure \ref{fig: sm_rate_hist}). The assembly rate of BCGs reveals a more heterogeneous history, with several secondary peaks suggesting that many of these objects experienced multiple merger events, especially in more massive halos.

The stars that will ultimately reside in BCGs are formed earlier than those in other massive galaxies, reaching 50\% of their mass by $z = 3.5 - 4$ (Figure \ref{fig: sm_form_hist}). The total stellar mass of all BCG progenitors exceeds the final BCG mass at $\rm z \lesssim 2.5$, as stars are stripped from the progenitor galaxies when they approach the central region, becoming part of the intracluster light (ICL) or the extended stellar halo of the BCG. 

At $\rm z \sim 0.1$, over $90\%$ of the stellar mass of BCGs originated from other galaxies that merged with their main progenitor (Figure \ref{fig: sm_origin}), a proportion higher than that observed for other massive galaxies. Additionally, BCGs become bulge-dominated earlier than other galaxy types. By around $\rm z \sim 3$, nearly all protoBCGs are bulge-dominated (Figure \ref{fig: sm_bulge}).


Finally, (proto)BCGs become the most massive galaxy in their structures relatively late, around $\rm z \sim 1.3$, roughly the redshift at which the structures are transitioning from protocluster to cluster. By $\rm z \sim 0.1$, BCGs are approximately 1.3 to 3.5 times more massive than the second most massive galaxy in the structure (Figure \ref{fig: sm_ratio}). This behavior is explained by the dense environment in which (proto)BCGs are embedded—close to the bottom of the gravitational potential field (Figure \ref{fig: offset})—with a consistently higher density in both galaxy number and stellar mass throughout their history (Figure \ref{fig: density_ratio}).

Given the results presented here, we can conclude that BCGs indeed have a distinct history compared to other massive galaxies. This distinction is directly linked to the mass-rich environments their progenitors inhabited throughout their history, positioning them in a uniquely advantageous location to grow their mass more than any other galaxy in the universe.

In future work, we aim to extend this analysis by investigating how the star formation rate of these galaxies evolves as a function of stellar mass over time. We also intend to explore how the fraction of quiescent galaxies relates to the different galaxy types, in order to identify the dominant processes—both environmental and internal—that drive galaxy quenching. Finally, our results can be tested observationally with existing cluster and protocluster catalogs. The main challenge in this comparison is the difference in how protoclusters are defined in simulations and in observations. A possible step forward is the development of a quantitative measure of the degree of virialization of structures, which could provide a more consistent framework for connecting simulated and observed systems.


\section*{Acknowledgements}
MCV acknowledges the Fundaç\~ao de Amparo à Pesquisa do Estado de S\~ao Paulo (FAPESP; 2021/06590-0) for supporting his PhD and Research Internship Abroad at the Department of Astrophysical Sciences, Princeton University. He also thanks the Department of Astrophysical Sciences at Princeton University for its financial support in making this internship possible. LSJ acknowledges the support from CNPq (308994/2021-3) and FAPESP (2011/51680-6). PA-A thanks the Coordenaç\~ao de Aperfeiçoamento de Pessoal de Nível Superior – Brasil (CAPES), for supporting his PhD scholarship (project 88882.332909/2020-01). DS is funded by the Spanish Ministry of Universities and the European Next Generation Fond under the \textit{Margarita Salas Fellowship} CA1/RSUE/2021-00720 and acknowledges the support from the European Union’s HORIZON-MSCA-2021-SE-01 Research and Innovation Programme under the \textit{Marie Sklodowska-Curie} grant agreement number 101086388 (LACEGAL). She wishes to thank \textit{La Sra.\ Pop}, whose soulful spirit and generous stage brought art to life and turned ideas into shared experience.

\vspace{5mm}

\software{Numpy \citep{harris2020}, Pandas \citep{reback2020}, Scipy \citep{scipy}, Matplotlib \citep{hunter2007}, Astropy \citep{astropy}, Sklearn \citep{pedregosa2011}}.




\bibliography{bibliography}{}
\bibliographystyle{aasjournal}

\end{document}